\documentclass[twocolumn,english,longbibliography,nofootinbib]{revtex4-1}
\usepackage[T1]{fontenc}
\usepackage[utf8]{inputenc}
\usepackage{geometry}
\usepackage{array}
\usepackage{verbatim}
\usepackage{amsmath}
\usepackage{amssymb}
\usepackage{graphicx}
\PassOptionsToPackage{normalem}{ulem}
\usepackage{ulem}

\makeatletter

\providecommand{\tabularnewline}{\\}

\usepackage[colorlinks]{hyperref}

\makeatother

\usepackage{babel}
\begin{document}
\title{Inter-Party Avalanche Involvements May Increase Quadratically With
Party Density}
\author{Charles A. Hagedorn}
\address{Center for Experimental Nuclear Physics and Astrophysics, Box 354290,
University of Washington, Seattle, Washington 98195-4290}
\email{cah49@uw.edu}

\begin{abstract}
We estimate, from first-principles, the rate of inter-party avalanche
involvements. The model suggests that the likelihood of inter-party
involvements is quadratic in the density of parties – twice as many
parties quadruples the likelihood. The model predicts that when the
product of the party-density and the area of a day's potential avalanches
approaches one, inter-party avalanche involvements will become a substantial
fraction of all avalanche involvements. As a corollary, the relative
rate of inter-party involvements is expected to increase with avalanche
size. We argue, with selected North American inter-party incidents
from 2001-2019, that inter-party involvements are a timely concern.
To spur conversation, we enumerate a variety of strategies that may
mitigate inter-party hazard.
\end{abstract}
\maketitle
\begin{center}
\emph{For Monty Busbee\citep{KendallDecember19}, and for you.}
\par\end{center}

\section{Introduction}

The number and density of backcountry travelers continues to increase\citep{WWAHumans,AvalancheHourEvelynLees,MeisenheimerNSAW,BackcountryCrowdingDynamics}
in the United States and around the world. Many have written\citep{BackcountryMagazineCrowding,BackcountryComPowderPolice,SlidePodcastScene,CrestedButteEtiquette}
about the emerging hazard of inter-party involvements. At some density
of backcountry travelers, avalanches triggered by one party will be
likely to strike another. This work gives formal voice to this idea
and, further, suggests that the time when such involvements become
frequent is either now or in the near future. In Section \ref{sec:Estimation},
we estimate the relevant critical density from first principles. In
Section \ref{sec:Significant-Accidents}, we enumerate several recent
inter-party avalanche incidents in North America. In Section \ref{sec:Mitigation},
we enumerate and discuss strategies that may partially mitigate the
risk.

This work provides a framework within which to discuss and estimate
the magnitude of the concern, addresses it directly as a hazard, and
attempts to spur discussion of strategies to reduce the likelihood
of future inter-party involvements.

\section{Estimation\label{sec:Estimation}}

To assess inter-party avalanche involvements as a concern, it is useful
to develop a first-principles model for their likelihood. The estimation
that follows is a simple approach that is likely to require small
corrections to yield quantitative predictions. This model has qualitative
predictive power, predicting both that the inter-party involvement
rate is quadratic in party density and that there is a critical party-density
at which the inter-party involvement rate becomes significant.

\begin{figure}
\begin{centering}
\includegraphics[width=1\columnwidth]{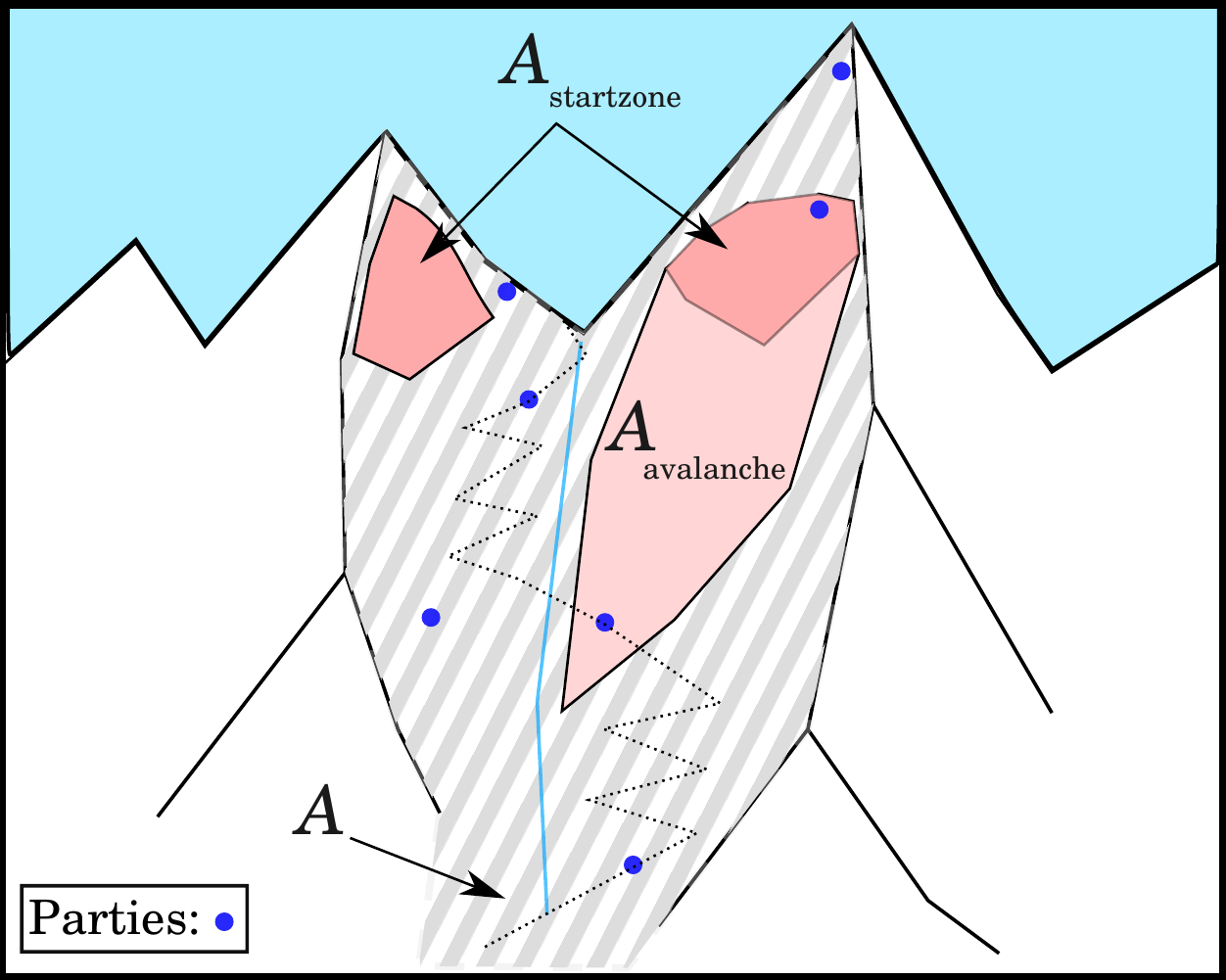}
\par\end{centering}
\caption{Diagram showing the definitions of $A$, $A_{\text{startzone}},$and
$A_{\text{avalanche}}$. $A_{\text{startzone}}$ is the area of a
day's typical individual start-zone. $N_{\text{parties}}$ here is
7.\label{fig:Diagram-showing-the}}
\end{figure}

\subsection{Derivation\label{subsec:Derivation}}

Suppose that there are $N_{\text{parties}}$ parties of winter travelers
in a region of area $A$, shown in Figure \ref{fig:Diagram-showing-the},
where $N_{\text{parties}}>1$. This analysis is independent of the
definition of a ``party'', but for the estimates that follow, we
define a party as ``any group, as small as one person, that travels
independently''. Let $n_{\text{party}}=N_{\text{parties}}/A$ be
the party-density. Here, we define an avalanche ``involvement''
as an interaction between a party and an avalanche, an ``incident''
as an event in which at least one involvement occurs, and an ``accident''
as an incident that yielded significant injury or death. If there
are few-enough parties that they are only ever caught in their own
avalanches, one expects the avalanche-involvement (and incident) rate
per unit area per unit time, $r_{\text{single-party}}$, to be proportional
to the party density:

\begin{equation}
r_{\text{single-party}}=r_{0}n_{\text{party}}\label{eq:Single party avalanche rate}
\end{equation}

The implication is straightforward: More parties, proportionally more
involvements. We estimate the many factors subsumed into $r_{0}$
(avalanches per party per unit time) in Appendix \ref{subsec:Deriving-likelihood}.

What if there are enough parties that there is a chance one group
could trigger a slide that hits another group?

There are two important factors:
\begin{enumerate}
\item When there are more parties in an area, they will trigger more avalanches
(rate $r_{0}n_{\text{party}}$), as before. 
\item If the average avalanche that day sweeps through an area $A_{\text{avalanche}}$,
the number of other parties that will be struck by that avalanche
is approximately\footnote{We assume a uniform distribution of parties and that $N_{\text{parties}}$
is ``large''. The result for any $N_{\text{parties }}$ is derived
in Appendix \ref{sec:Correctly-handling-N}. The correction to Equation
\ref{eq:Quadratic Scaling} is $(N_{\text{parties}}-1)/N_{\text{parties}}$,
never more-impactful than a factor of two.} $A_{\text{avalanche}}n_{\text{party}}$.
\end{enumerate}
So, the rate of inter-party avalanche involvements, $r_{\text{interparty}}$,
ought to scale like

\[
r_{\text{interparty}}=r_{\text{single-party}}A_{\text{avalanche}}n_{\text{party}}
\]

Thus,

\begin{equation}
r_{\text{interparty}}=r_{0}A_{\text{avalanche}}n_{\text{party}}^{2}.\label{eq:Quadratic Scaling}
\end{equation}
 \textbf{The rate of inter-party involvements should scale like the
party-density }\textbf{\emph{squared}} ($n^{2}$). This is the key
observation.

Why is $n_{\text{party}}^{2}$ important? Let's look at the rate that
parties will have an involvement of either kind by summing the rates:

\[
r_{\text{incident}}=r_{\text{avalanche}}+r_{\text{interparty}}
\]

That is:

\[
r_{\text{incident}}=r_{0}(n_{\text{party}}+A_{\text{avalanche}}n_{\text{party}}^{2})
\]

or, suggestively,

\begin{equation}
r_{\text{incident}}=r_{0}n_{\text{party}}(1+A_{\text{avalanche}}n_{\text{party}})\label{eq:Combined Single and multi-party accident rate}
\end{equation}

\begin{figure}
\begin{centering}
\includegraphics[width=1\columnwidth]{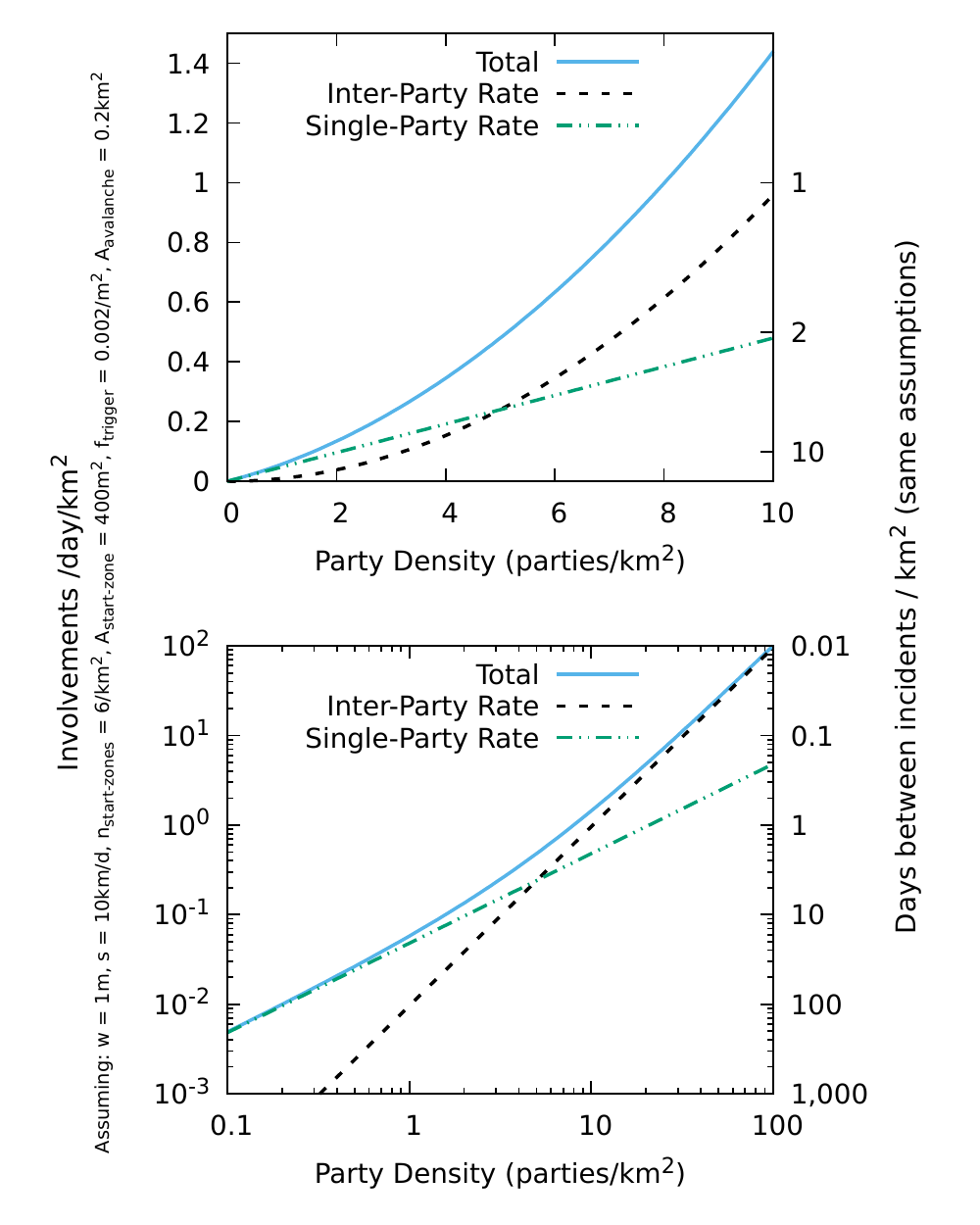}
\par\end{centering}
\caption{Illustrative plots showing quadratic growth in inter-party involvements
surpassing linear growth in single-party involvements when $n_{\text{party}}A_{\text{avalanche}}$
exceeds 1; when there is more than one party per avalanche-area. The
inter-party critical density for an $A_{\text{avalanche}}=0.2$~km$^{2}$
avalanche, as in this example, is $n_{\text{critical}}=5$~parties/km$^{2}$.
The vertical scale will vary widely with snowpack properties (see
Appendix \ref{subsec:Deriving-likelihood}), but the relative rates
between single-party and inter-party involvements depend only on $A_{\text{avalanche}}$.
The parameters chosen for these plots highlight the importance of
both avalanche size and party density – when avalanches are large
or parties are close together, inter-party involvement becomes likely.\label{fig:Illustrative-example-of}}

\end{figure}

An illustrative plot of these rates is shown in Figure~\ref{fig:Illustrative-example-of}.

If the party density, $n_{\text{party}}$, is small, then $n_{\text{party}}^{2}$
is smaller still, just as $(0.1)^{2}$ is 0.01. If $n_{\text{party}}$
is larger than 1, though, then $n_{\text{party}}^{2}$ dominates the
problem, just as $10^{2}$ is 100. What, then, sets the scale of \textquotedbl 1\textquotedbl{}
?

Equation \ref{eq:Combined Single and multi-party accident rate} holds
the answer: Once $A_{\text{avalanche}}n_{\text{party}}\gtrsim1$,
inter-party involvements will dominate. This makes sense – once there
is roughly one party per avalanche-area, any triggered avalanche is
likely to hit another party. This notion is intuitive – in-bounds
avalanches at ski areas are scary in large part because the skier-density
is so high. 

We define this critical density as $n_{\text{critical}}\equiv1/A_{\text{avalanche}}$.
It is of particular utility because it depends \uline{only} on
the expected size of a day's avalanches, and not on any other property
of a day's anticipated instability, including the likelihood of triggering
a slide. 

While it is difficult to predict the absolute involvement rate, as
the determination of $r_{0}$ is uncertain, one can consider the ratio
of inter-party involvements to total involvements:

\begin{align}
\frac{r_{\text{interparty}}}{r_{\text{incident}}} & =\frac{A_{\text{avalanche}}n_{\text{party}}^{2}}{n_{\text{party}}+A_{\text{avalanche}}n_{\text{party}}^{2}}\label{eq:RelativeIncidentFraction}\\
 & =\frac{1}{\frac{1}{n_{\text{party}}A_{\text{avalanche }}}+1}\nonumber 
\end{align}

\begin{figure}
\begin{centering}
\includegraphics[width=1\columnwidth]{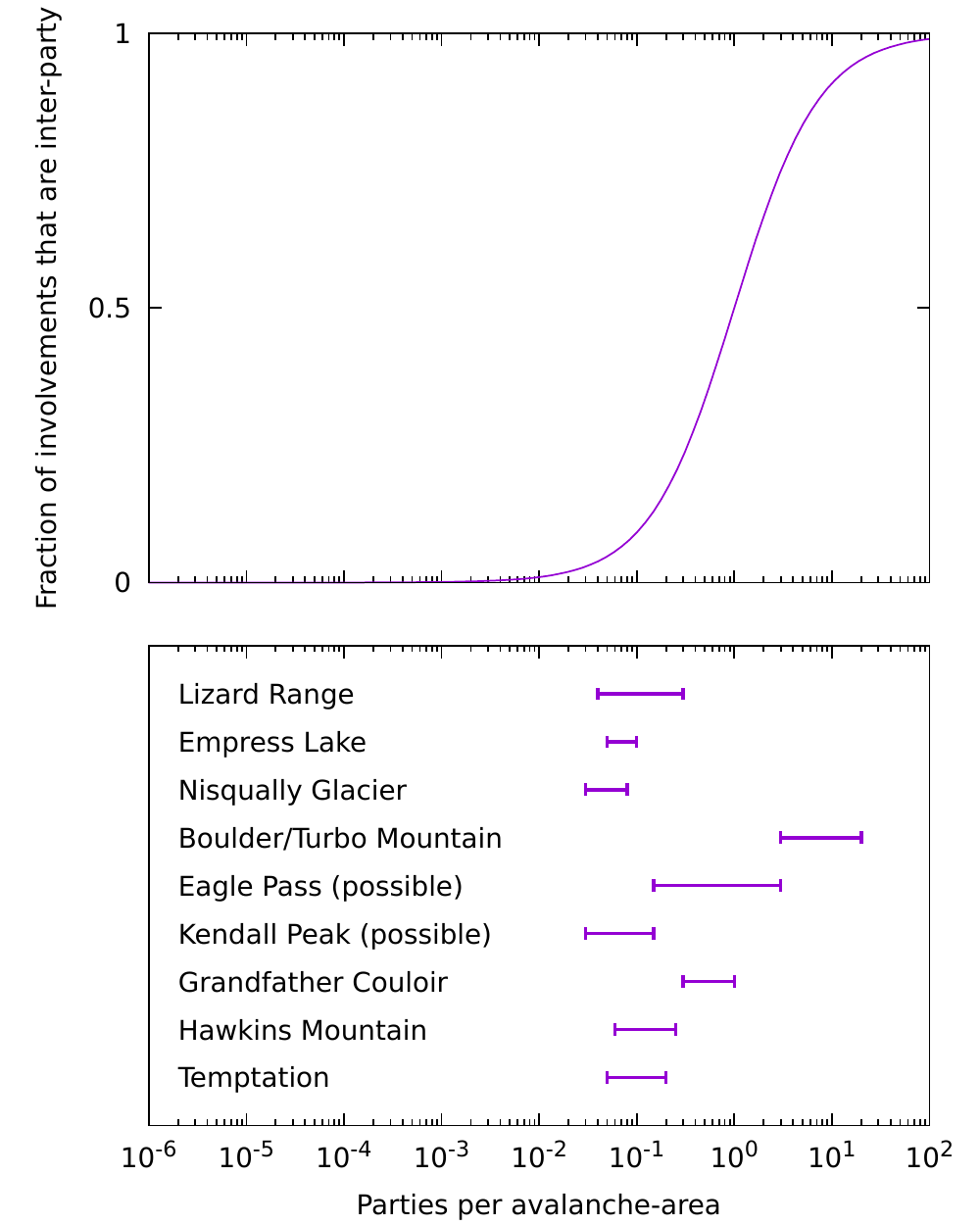}
\par\end{centering}
\caption{Upper Panel: Approximate fraction of inter-party avalanche involvements
as a function of party-density, measured in units of avalanche-area,
as described by Equation \ref{eq:RelativeIncidentFraction}. Lower
Panel: Approximate values of $n_{\text{party}}A_{\text{avalanche }}$
from incidents described in Section \ref{sec:Significant-Accidents}.
It is interesting that all but three of the incidents occurred with
$n_{\text{party}}A_{\text{avalanche}}$ near 0.1.\label{fig:Approximate-fraction-of}}
\end{figure}

As shown in Figure \ref{fig:Approximate-fraction-of}, this has the
expected behavior, where inter-party involvements become significant
as $n_{\text{party}}A_{\text{avalanche}}$ approaches 1.

\subsection{Relationship of inter-party involvement rate to avalanche size}

Equations \ref{eq:Quadratic Scaling} and \ref{eq:Combined Single and multi-party accident rate}
have an important corollary: If the size or likelihood of a day's
avalanches is greater, we should expect the rate at which inter-party
involvements occur to be greater. If slides are more likely, $r_{0}$
will tend to increase, and if slides are larger, $A_{\text{avalanche}}$
is larger. As backcountry travelers, larger and more-likely slides
will force wary travelers to consider larger areas (further uphill)
and more places (increased number of potential start zones) in the
context of inter-party incidents. In our model, incidents with $A_{\text{avalanche}}n_{\text{party}}\gtrsim1$
are likely to include inter-party involvements. Indeed, if $A_{\text{avalanche}}$
is large-enough, as in the 0.13~km$^{2}$ Jumbo Mountain slide\citep{JumboMountain},
backcountry travelers can even bury parties sitting at home on a sofa.

\subsection{Relationship of inter-party involvement rate to terrain}

Looking at the equations alone, one would expect both the incident
rate and the inter-party involvement rate to be relatively low. Even
in mountainous terrain, start-zones are comparatively small, and much
of the area is not avalanche terrain.

Critically, however, \emph{backcountry travelers} are often concentrated
in both start zones and avalanche terrain by the desire for quality
skiing and the travel-focusing effects of terrain itself. Furthermore,
limited access points (mountain passes/trailheads) for those parties
tend to further concentrate those parties into yet higher densities. 

These concentrated party densities can be large. As an example, consider
``The Slot'', a popular backcountry ski route at Snoqualmie Pass,
WA. In its narrow upper reaches, it is perhaps 30~m wide and 250~m
long. If there are but two parties in the couloir, the effective party
density is greater than 250~parties/km$^{2}$. For scale, the population
density of the nearby city of Seattle, Washington is $\sim3400$ people/km$^{2}$
\citep{CensusDensity}.

\subsection{Weather and snow conditions as density and consequence amplifiers}

\subsubsection{Storm-skiing}

It is a guess, but the Pacific-Northwestern inter-party avalanche
hazard may be at its greatest on deep-snow storm-skiing days, when
trail-breaking is arduous, avalanches are likely, and powder fever
is high. Arduous trail-breaking may be the most-powerful concentrator
of backcountry travelers, amplifying the already-powerful magnetism
of an extant skintrack.

On storm-skiing days, day-tripping backcountry travelers embark from
a limited set of trailheads and head for a limited set of ``safe''
objectives. In Western Washington State, with a population of thousands
of active backcountry travelers\citep{NWACAnnualReport2018}, there
are approximately nine such popular trailheads. As skintracks are
forged more-deeply into the morning, they bifurcate, but the number
of traditional and natural ski options remains slim.

\subsubsection{Optimal conditions}

``Good'' conditions can concentrate parties. Mountains and routes
offer only certain optimal times for passage. In the Cascades, and
perhaps everywhere, this can be seen prominently in periods of spring/summer
skiing; both snowpack and limited access curtail skiing at other times.
There are narrow optimal windows for all of Tahoma/Rainier's Fuhrer
Finger, the North Face of the Northwest Ridge of Pahto/Adams, and
Ulrich's Couloir of Stuart – these are listed for their combination
of popularity and constrained terrain, where a small summer slide
can ravage anyone below. Indeed, on April 1, 2019, it was reported\citep{HesserTetongBackup}
that 16 people attempted, and bottlenecked, the ski route on the Grand
Teton.

Wintertime weather windows provide perhaps even more temporal concentration,
as the number of good days for attempting bold routes is slim, perhaps
only one or two weekends per season. Fortunately, for now, the number
of skiers attempting bold winter routes seems low-enough that inter-party
hazard is likely to be a sub-dominant concern.

In addition to temporal concentration at the scale of days, safety
and snow-quality considerations may also concentrate parties on shorter
durations. Springtime daily melt-freeze/corn cycles can constrain
safe travel to time intervals measured in hours. The arrival of a
storm or imminent warming can force travelers to enter or cross slopes
before hazard increases. Finally, the perceived pressure from other
people, whether to make the first tracks on a slope or, again paradoxically,
to cross/descend a slope \emph{before} the presence of other parties
can make the slope more hazardous, can concentrate parties in small
areas.

\subsubsection{Reduced visibility}

If nearby parties cannot be seen, as in the Lizard Range incident
(Section \ref{subsec:Lizard-Range}), and the distances between parties
are greater than that allowed by voice contact, it is impossible to
be aware of anyone above or below. In low-visibility conditions in
densely-traveled areas, this may drive us to avoid descending or ascending
an otherwise-reasonable route simply because it may not be possible
to verify that the route is free of other travelers.

\section{Significant Incidents\label{sec:Significant-Accidents}}

Here, we enumerate a series of inter-party incidents, near-misses,
and possible inter-party incidents. We provide a subjective summary
from an inter-party-involvement perspective. The references offer
expanded detail. Several of the incidents appear to have been reported
only informally. It is a surprise that it has been difficult to find
formal accident reports for some modern Canadian avalanches, each
of which have required rescue or a coroner.

While the catalyst for this paper is the accident described in Section
\ref{subsec:December19}, these incidents are presented in chronological
order. This enumeration is not intended to be comprehensive, but rather
to provide both a sense of the varied types of inter-party involvements
and to make clear that these involvements happen.

For incidents where documentation makes it possible, we have attempted
to measure the areas in question, the number of parties, and compute
both the party density and $n_{\text{party}}A_{\text{avalanche}}$,
summarized in Table \ref{tab:Quantitative-characterization-of}. The
areas were measured using Caltopo\citep{Caltopo} and attempt to cover
the terrain that would reasonably be regarded as ``connected'' by
an observer on the ground; this is a somewhat-subjective quantity
(see Appendix \ref{subsec:Sensitivity-to-area}). At Krause's suggestion\citep{DougKrauseEmail},
we include the avalanche type, to the extent that it is reported in
the record.

\subsection{Lizard Range, February 13, 2001\label{subsec:Lizard-Range}}

A party of five skiers crossed over a ridgeline outside the Fernie
Ski Area, intent on skiing a constrained 35-40-degree route. As they
prepared for descent, a ski cut triggered a 24~m-wide and 28~cm-deep
wind-slab avalanche\citep{CanadaAvalancheBook,LizardRangeAvalancheCanada,LizardRangeCSAC,LizardRangeGlobeAndMail,LizardRangeOttawa},
sweeping down the route and out of sight. At approximately the same
time, a party of thirteen foreign skiers was hit by a size-2.5 avalanche
while traversing the side of the valley below. Six skiers were caught,
resulting in two fatalities. A third party witnessed the entire incident
and immediately returned toward the ski area to raise an alarm. On
the way, they encountered the first party, which was unaware of the
slide's impact on the second party.

\subsection{Empress Lake, March 20, 2004}

A party of snowmobilers\citep{CanadaAvalancheBook,EmpressLakeAvalancheCanada,HankeNarrative1,Hanke2,EmpressGlobeAndMail,EmpressCastanet}
assessed a cliff-drop in steep terrain, concluding that the expected
instability was sufficiently manageable. The cliff-drop was attempted,
triggering a small slab and apparently damaging a sled. As the party
regrouped mid-path and assessed the damage, a second party appeared.
The second party climbed immediately into the now-overhead start-zone,
triggering a D2.5 slide ($30$~m$\times125\text{-}150$~m$\times150$~cm
crown). Three were caught, one was killed.

\subsection{Nisqually/Wilson Glaciers, June 14, 2008}

On a weekend that surprised skiers on at least three Cascadian volcanoes\citep{CharlieWyeast,CrokattHood,HummelCircumnav,CharlieSWChutes}
with large summer slab avalanches, a party of skiers descending the
slopes of Tahoma/Mount Rainier triggered a slab\citep{RyanWilsonThread}
that caught an ascending solo splitboarder and nearly caught an ascending
climbing party at the convergence of the Wilson and Nisqually glaciers.
The splitboarder and triggering skier were carried and sustained only
minor injury. A second trip-report\citep{HelmstaderWilson} from Rainier's
south side emphasizes the day's instability during what is generally
prime volcano-ski season.

\subsection{Boulder/Turbo Mountain, March 13, 2010 \label{subsec:Turbo}}

Perhaps the most-spectacular inter-party accident\citep{CodeBlackBoulderMountain,BoulderMountainResponseReview,TurboHillPreliminaryReport}
of which the author is aware: A snowmobiling festival/competition,
with snowmobiles highmarking above $\sim200$ spectators at the time
of the slide, was directly struck by a D3 slide presumed to have been
triggered by a participant. Approximately forty people were buried,
with 32 injured and two fatalities. The on-scene coordination and
rapid rescue response were remarkably effective for such a chaotic
event.

After-action review\citep{ParksCanadaBoulderMountainPresentation}
notes the probability of legal charges in similar future events.

\subsection{Eagle Pass Avalanche, March 19, 2010}

Only an Incident Summary\citep{EaglePassAvalancheCanada}, an Avalanche
Involvement Report\citep{EaglePassCSAC}, and an associated photograph
appear to be available for this accident; this synopsis is composed
from news articles and indirect references in the Boulder Mountain
accident reporting. One week following the Boulder Mountain slide,
a snowmobiling party of two may have triggered a D3.5 slide above
one or two parties of at least ten snowmobilers, yielding one fatality
and at least one injury\citep{EaglePassNewsArticle1,EaglePassNewsArticle2,CodeBlackBoulderMountain}.
The total number of people involved may have been as large as 21\citep{ParksCanadaBoulderMountainPresentation}.

\subsection{Little Cottonwood ``Frenzy'', November 13, 2011}

In November of 2011, the year's first skiable snow arrived atop faceted
autumn snow, drawing many backcountry skiers into the un-opened Little
Cottonwood Canyon ski resorts. At least 11 human-triggered avalanches
were reported, yielding one fatality and one broken femur. Perhaps
most-striking in the fatality's accident report\citep{GadValleyUACFrenzy}:
\begin{quote}
It was reported that other parties at Alta continued to ski and knock
down avalanches into Greeley Bowl while the rescue was in progress.
Creating another incident during this situation is unacceptable.
\end{quote}

\subsection{Taylor Mountain Near-Miss, January 24, 2012}

A skier ski-cutting a popular run on a reportedly-unstable day triggered
a D3.5, R4 avalanche\citep{TaylorMountainJHAvalanche} that crossed
the uptrack for nearby slopes and ascended an adjoining slope. The
uptrack was buried under up to twenty feet of debris. Fortunately,
nobody was caught, though a small shift in timing would have yielded
an inter-party accident. The outcry and subsequent discussion\citep{TaylorMountainTARMusings,TaylorMountainTetonAT,TaylorMountainTetonValleyNews}
was intense and sustained.

\subsection{Kendall Peak, December 19, 2015\label{subsec:December19}}

This accident is the catalyst for this work. As detailed by the NWAC
accident report\citep{KendallDecember19}, a solo skier disappeared
after early afternoon on December 19, the first deep storm-skiing
Saturday of the 2015/2016 season at Snoqualmie Pass\citep{KendallTrap}.
The number of skiers and parties in the area was unusually high, with
at least nine parties recreating in a single square kilometer. In
Table \ref{tab:Quantitative-characterization-of}, the estimations
of $A$ and $N$ are correlated, hence the diminished range of $n_{\text{party}}$.

After an extensive search effort, exceeding 3000 person-hours, the
solo-skier's remains were recovered on June 4, 2016. The accident
location, and the injuries to the skier, were consistent with an avalanche-related
fatality. Subsequent investigation revealed that two parties had triggered
avalanches uphill of the accident site on the afternoon of the skier's
disappearance. Furthermore, the first of those avalanches can be regarded
as an inter-party near miss, as a minute's difference in timing would
have seen the second party struck by the first. NWAC's analysis of
the accident does not take a position on whether the fatality was
caused by an inter-party avalanche – both a third avalanche triggered
by the soloist or a natural avalanche remain viable hypotheses.

\subsection{Rogers Pass, Avalanche Crest, February 14, 2016}

A party of two led the way into an alpine zone in an attempt to get
ahead of the Rogers Pass crowds\citep{AvalancheCrestBigLines,AvalancheCrestRogersPassScared}.
After a short lap, they encountered four other parties (from four
nationalities!) on their way up Avalanche Crest proper. Two of the
foreign parties trailed them as they pushed higher. The party of two
reached their highpoint and skied. Before they were clear of the line,
one of the parties above began to ski, triggering a D2.5-3 avalanche\citep{AvalancheCrestAvalancheVideo}.
Both of the skiers below were caught and carried, one sustaining life-threatening
injury.

\subsection{Grandfather Couloir, April 3, 2016\label{subsec:Grandfather-Couloir}}

As detailed by the CAIC accident report\citep{GrandfatherCouloirTwoPartyInteraction}
and subsequent lawsuit\citep{SanMiguelCountyJudgement,GrandfatherLegalBlog},
two parties of two met above a consequential couloir with a mandatory
mid-route rappel. The couloir is adjoined by Oblivion Bowl, which
also funnels through the mid-route choke. Recognizing the inter-party
hazard, the two parties agreed to let Group 1 ski and rappel first,
while Group 2 waited above for a cell-phone call. (Group 1 had radios,
and the region, Bear Creek, has a pioneering community radio protocol,
see Section \ref{subsec:Radios}, but Group 2 did not.) Group 1 skied
the couloir without incident but, while mid-rappel, was struck by
a slide from above, sustaining injury. Group 2 had waited roughly
45 minutes and attempted to call Group 1 once. Warming snow and the
imminent arrival of at least one more Group had spurred Group 2 into
skiing before receiving a call from Group 1. Group 2 triggered an
R1D1 sluff – the same slide that struck Group 1.

The judgment from the San Miguel County Court is notable for its detailed
analysis of the accident, its humanity, and reasoned approach to assigning
fault. Pursuant to important details in the case, Group 2 was found
to be financially liable for the injuries sustained by Group 1.

\subsection{Mt. Herman, March 4, 2017\label{subsec:Herman}}

As reported in NWAC's accident summaries\citep{1617AccidentSummaries}:
\begin{quote}
Widespread 1-2 ft storm slabs and larger 3-5 ft wind slabs were reported
in the backcountry near Mt Baker on Saturday, March 4th. An incident
occurred on Mt Herman when a large wind slab on an east aspect was
triggered from a party above, partially burying two and completely
burying one in a separate party at the base of the slide path. The
impacted party was transitioning back to climbing skins when they
were caught in the avalanche.
\end{quote}

\subsection{Hawkins Mountain, March 4, 2017 \label{subsec:Hawkins}}

As detailed by the NWAC accident report\citep{HawkinsMountain}, a
party of two snowmobilers triggered a large slide (D2+). At the time
of the slide, a separate party of two snowmobilers was eating lunch
in the runout zone, with their machines turned off. The lower party
ran from the slide – their sleds and rescue gear were buried. The
lower party was able to rescue one of the triggering snowmobilers
and locate/extract the body of the second\citep{SnowestHawkins}.

\subsection{Bear Creek, Temptation slide path, February 19, 2019}

A group of three snowboarders left the Telluride ski area to ski a
permanently-closed but frequently-skied\citep{TemptationSlideArticle}
route. Partway through the descent, a boarder triggered a slide that
stepped down, D2 in size, running 2,000 vertical feet\citep{TemptationCAICAccidentReport}.
The slide buried six feet deep a solo skier out for a short exercise
jaunt on a heavily-used trail.

Seeing many tracks entering and leaving the deposition zone, the uphill
party beacon-searched and spot-probed the debris pile without result
before returning to town. The solo skier, reported missing by his
wife, was located by probe-line the following day.

\begin{table*}
\begin{centering}
\begin{tabular}{|c|>{\centering}p{2.6cm}|c|c|>{\centering}p{2.8cm}|c|c|}
\hline 
Incident & Type & $A$ (km$^{2}$) & $N_{\text{parties}}$ & $n_{\text{party}}$(parties/km$^{2}$) & $A_{\text{avalanche}}$(km$^{2}$) & $n_{\text{party}}A_{\text{avalanche}}$\tabularnewline
\hline 
\hline 
Lizard Range & Wind Slab,\linebreak{}
cross-loading & 2-4.5 & 3+ & 0.7-1.5 & $\sim0.1$ & 0.04-0.3\tabularnewline
\hline 
Empress Lake & Dry Slab & $\sim1.2$ & 2 & 1.6 & $\sim0.05$ & 0.05-0.1\tabularnewline
\hline 
Nisqually/Wilson & Slab & $\sim1.3$ & 3+ & $\geq2.3$ & $\sim$0.02 & 0.03-0.08\tabularnewline
\hline 
Boulder/Turbo Mountain & Persistent Slab & 1-2 & 30-100 & 15-100 & $\sim0.2$ & 3-20\tabularnewline
\hline 
Eagle Pass (possible) & Slab & 1-2 & 2-3 & 0.5-3 & $>0.3$ & 0.15-3\tabularnewline
\hline 
Taylor Mountain & Hard Slab & $\sim0.8$ &  &  & $>0.3$ & \tabularnewline
\hline 
Kendall Peak (possible) & Slab & 0.2-1.0 & 3-9 & 9-15 & 0.003-0.01 & 0.03-0.15\tabularnewline
\hline 
Avalanche Crest/Rogers & Slab & 2-5 & 5+ & 1-5 &  & \tabularnewline
\hline 
Grandfather Couloir & Loose Snow & $\sim0.16$ & 2 & $\sim12.5$ & 0.02-0.08 & 0.3-1\tabularnewline
\hline 
Mount Herman & Wind Slab & 0.4-1.2 & 2+ & $>2\text{-}5$ &  & \tabularnewline
\hline 
Hawkins Mountain & Soft Slab & 0.6-1 & 2-3 & 2-5 & 0.03-0.05 & 0.06-0.25\tabularnewline
\hline 
Bear Creek, Temptation & Soft Slab & $0.3$-$1.3$ & 2+ & 1.5-7 & $\sim0.03$ & 0.05-0.2\tabularnewline
\hline 
\end{tabular}
\par\end{centering}
\caption{Quantitative characterization of inter-party incidents. The estimation
of $A$ is inherently subjective and uncertain. The model predicts
that inter-party incidents become likely as $n_{\text{party}}A_{\text{avalanche}}$
approaches 1. Eleven out of twelve incidents involve slab avalanches.
\label{tab:Quantitative-characterization-of}}
\end{table*}

\subsection{Common threads}

While this enumeration is not intended to be comprehensive, it is
interesting to look for commonalities among the incidents. Two stand
out: 1) For those incidents where $n_{\text{party}}A_{\text{avalanche}}$
can be estimated, 6 out of 9 overlapped $n_{\text{party}}A_{\text{avalanche}}\sim0.1$.
This is smaller than the most-natural expectation of Equation~\ref{eq:RelativeIncidentFraction},
but it is encouraging to find a possible invariant. The model is,
of course, blind to any propensity for parties to avoid one-another.
2) Eleven out of twelve incidents involved slab avalanches. Whether
this is simple correlation with size/frequency or more meaningful
must await further study. The Grandfather Couloir incident shows that
loose slides can cause inter-party incidents as well\footnote{It is the author's opinion that there will eventually be a loose-wet
inter-party incident in the Cascades, as the popularity of accessible
Spring skiing continues to grow.}.

\section{Mitigation\label{sec:Mitigation}}

It is unlikely, and perhaps undesirable, that the number of backcountry
travelers will be reduced. Therefore, we must find ways to limit the
likelihood of inter-party incidents. The following are strategies
that may mitigate the risk. There are certainly more – these are suggested
to organize and catalyze discussion.

\subsection{Awareness and Education}

Awareness of the increasing likelihood of inter-party conflict with
density may, on its own, help to curtail inter-party involvements.
That is, of course, the intent of this work. In places where the density
of travelers approaches the critical density, the hazard should be
discussed and mitigated. At least a passing mention of inter-party
hazard in avalanche education at all levels now seems prudent.

As better understanding of the factors that lead to inter-party avalanche
involvements becomes understood, it will make sense to bring light
to the hazard in avalanche bulletins. A key takeaway for forecasters
is the apparent importance of avalanche size in determining the likelihood
of inter-party involvements, especially in situations (\emph{e.g.}
persistent weaknesses) where avalanche size alone may not convince
travelers to stay home.

\subsection{Density reduction}

The root of inter-party incidents is not the \emph{number} of backcountry
travelers, but their \emph{density}. Backcountry travelers use a tiny
fraction of North American alpine terrain, perhaps primarily due to
access constraints. Following the Kendall Peak accident\citep{KendallDecember19},
the author has sharply curtailed trips into densely-skied terrain,
choosing lesser-known destinations, difficult access, or lower-quality
snow in order to decrease overhead risk from other parties. While
this strategy has limits, there are still lonely places with great
skiing to be found on the map.

As advocacy organizations work to expand wintertime access, it may
be reasonable to add safety to the reasons for improved access. Inferring
from our estimates, it is natural to encourage party densities below
one per characteristic-avalanche-size. D2 slides are roughly 0.01
km$^{2}$ in area. D3 slides are roughly 0.1 km$^{2}$ in area\citep{SWAG}.
Hence, encouraging party densities below 10-100 parties/km$^{2}$
\emph{in localized avalanche terrain} would be wise, and ten-times
lower (intimated by Figure \ref{fig:Approximate-fraction-of}) may
be prudent.

\subsection{Travel practices}

If density cannot be wholly avoided, then we must find ways to limit
our susceptibility to other parties and the risks we may pose to others.
We cannot control others' choices, but we can control our own.

\subsubsection{Defensive routefinding}

Like defensive driving, defensive routefinding consists of making
choices that mitigate the risks that others may pose. Careful selection
of routes through terrain, and, to a lesser extent, times to pass
through terrain, can strongly limit the chance that another party's
avalanche might hit us. 

Examples: 
\begin{itemize}
\item Many ridgelines are not susceptible to slides from above – traveling
upon such a ridgeline provides substantial protection from inter-party
hazard.
\item Some routes pass through areas of unavoidable avalanche hazard – there
are plenty of examples: The Mousetrap\citep{MousetrapAsulkanGuide}
at Rogers Pass, Source Lake Basin and Mushroom Couloir at Snoqualmie
Pass. Avoiding those areas during times that \emph{human-triggered}
slides are likely, rather than only when natural slides are likely,
will limit the risk of an inter-party incident.
\item The Hawkins Mountain (Section \ref{subsec:Hawkins}) and Turbo Hillclimb
(Section \ref{subsec:Turbo}) accidents make clear the danger of extended
loitering in inter-party-avalanche terrain. When assessing safe spots,
the likelihood of human triggers above must be considered.
\item Avalanches only happen in avalanche terrain – avoiding avalanche terrain
is a sure-fire way to avoid avalanche incidents of all kinds, including
inter-party incidents.
\end{itemize}
Choosing \emph{not} to travel (down or up) routes at times that we
expect others might be ascending or descending is also a form of defensive
routefinding. Anticipating the choices of others and making our own
conservative diversion or delay in order to limit the risks we might
present to others is a mark of a conscientious traveler.

\subsubsection{Active measures}

When we travel in start zones, we are potential triggers for avalanches.
In addition to passive measures to ensure that we are not traveling
above other people, there are small measures we can take to communicate
with those who might be below. Radios (see Section \ref{subsec:Radios})
are the superior method of communication in the backcountry, but are
not used by everyone, nor are they completely reliable. As a simple
measure, rock climbers and mountaineers yell ``Rope!'' before tossing
down a rappel rope, ``On Rappel!'' before rappelling (which can
dislodge rock), ``Off Rappel'' to communicate to those above and
below, and ``Rope'' again, when the rope is pulled (which can generate
more rockfall). The wording is important, but in the snow world, something
akin to ``Hello Below!'', ``Skiing!'', and ``Clear!'' could
provide warning and an opportunity for communication with those below.
Similarly, for a group entering constrained terrain from below, it
might be appropriate to shout (or radio (see \ref{subsec:Radios})),
``Entering Couloir!'' and make an occasional shout as they ascend.

Verbal warnings are, of course, antithetical to a low-impact and quiet
winter day (and may lead to confusion, as shouts are  a sign of an
emergency). If the rate of inter-party incidents continues to grow,
the necessity of such measures, and the approximate skier density
at which they are appropriate, will become self-evident. At a crowded
crag, it is perhaps more welcome than it is jarring to hear a climber
yell, ``Rope!'' as they begin an orderly descent.

In addition to verbal warnings, making micro-route decisions during
ascent and descent to maximize the leader's ability to see up/down-slope
can increase the likelihood of noting a potential inter-party conflict
before it becomes reality. 

\subsection{Regional travel standards for routes}

Just as we all agree to drive on one side of a road in order to prevent
head-on automotive collisions, community standards can improve mutually-beneficial
cooperation between parties who have never met.

\subsubsection{Run lists}

In recent years \citep{NWACRunList,RunListIsraelson}, the notion
of a ``Run List'' has moved from professional operations discussions
into the mainstream. A ``run list'' is a shared enumeration/mapping
of regions and ski runs used to facilitate discussion and improve
adherence to each day's terrain assessment/plan. In courses at the
Washington Alpine Club, run lists have become an integral part of
our courses, and our students often refer to them after the course
is complete. A publicly-shared run list can serve as a basis for conditions
reporting, incident response, and general communication. While run
lists themselves will not offer any advantage for inter-party incidents
– they may even serve to concentrate travelers during times of higher
hazard – a strong basis for communication is bedrock upon which many
mitigation strategies can be built.

\subsubsection{Terrain-specific traditions}

In certain high-density traffic locations, it may be helpful to establish
certain local traffic patterns. The aviation industry has ``Approach
Procedures'' and ``Instrument Approach Procedures'' (IAP)\citep{FAAApproaches}
for most airports that describe the airport-specific information needed
to land and take off safely. While IAPs contain a level of detail
not needed for backcountry travel, the occasional publicized traffic
pattern may be of use. These may be best described by examples:

Near the Mount Baker Ski Area, there is a backcountry ski run known
as ``Blueberry Chutes''. It is a high-traffic run in avalanche terrain,
1.5~km from a parking lot, with a straightforward and comparatively
safe alternative ascent route that reaches the run at the top. It
is also possible to ascend the run directly, which is arduous, slow,
exposed to avalanche hazard, and exposed to skiers dropping in from
above. In addition, those ascending from the bottom sometimes choose
to transition to downhill skiing mid-slope, spending even longer in
a hazardous spot. This location has become the site of inter-party
conflicts\citep{BlueberryBagleyTAY,BlueberryDecemberHop,BlueberrySeattleSkintrack}
and can deliver plenty of consequence\citep{BlueberryAprilFools,BlueberryAprilFoolsVideo}.
Such a location is a prime opportunity for a simple local understanding
– downhill traffic only in Blueberry Chutes – to be more-formally
known. While this strategy would have an impact on weaker skiers incapable
of skiing the steepest top pitch, it also protects those weaker skiers
from getting clobbered from above.

The Slot, a couloir on Snoqualmie Mountain near Snoqualmie Pass, WA,
was once regarded as much as a climbing route\citep{PhilEnigma} as
a ski line. In 2004, it was sufficiently rarely skied that it still
made occasional sense\citep{HummelSlot} to ascend directly up the
route for convenience. The Slot is now routinely skied, even by those
who ski before their day-jobs in Seattle. It is a rare thing to be
the first to the Slot's entrance after a storm, and it is common to
encounter multiple parties at the entrance. In the present environment,
it is hard to imagine a good reason to ascend the Slot in winter.
The hazard from above is significant.

The Alpental Ski Area sits in a constrained valley, with densely-used
ski-touring terrain at, and beyond, its head. The area has a relatively-unique
policy of limited avalanche control within its ``Back Bowls'' backcountry
terrain. The slide paths and runouts from that terrain reach the valley
bottom. Passing beneath the area during times of control work, or
traveling beneath the lift-accessible backcountry during times of
instability is inadvisable. Furthermore, the skiers returning to the
ski area are often traveling fast on the most-tempting skintrack.
Recent efforts\citep{NWACUphillPanel} have helped to separate uphill
and downhill traffic, and move uphill traffic to safer locations.
The US Ski Mountaineering Association maintains an updated list\citep{UphillPolicies}
of uphill travel and boundary policies for ski areas – it could make
sense to broaden it, or a similar repository, to include travel traditions
for high-traffic backcountry terrain.

\subsubsection{``Run board''}

One of the principal difficulties in avoiding conflict between disparate
groups is the simple awareness of the existence and location of other
parties. If parties leaving a trailhead were to note their number,
intended destination, and approximate schedule, perhaps akin to filing
a flight plan, subsequent parties would have the opportunity to know
who might be ahead, and where. Furthermore, if a radio channel were
appended, communication with those ahead might be possible.

The potential problems with such a system are myriad, but it may offer
utility to the conscientious.

A run board may be more useful in constrained terrain, like The Slot
at Snoqualmie Pass, where entering parties might note their number
and the time at which they entered, as a note for those who follow
behind. The 370~m-tall couloir has a visibility-obscuring dog-leg
about a third of the way down. Without coordination between parties
in the couloir, it is possible for a group descending the couloir
to be impacted by the actions of a group they have never seen. A conscientious
party arriving at the Slot's entrance could, seeing that a party had
entered the couloir only minutes before, choose to wait a few minutes
for those below to clear the run. The social-network share-ability
of a photo of a run board atop a popular run (\#firsttracks ?) might
increase the likelihood of its use. A ``run-clock'', fashioned after
the out-of-office ``back at 3pm'' clock-signs, altered to ``dropped
in at 7:45 am'' may suffice.

\subsection{Radios\label{subsec:Radios}}

Radios have emerged as an increasingly common tool in the backcountry
– remarkable for a sport which eschews complication and weight. In
the author's North American experience, FRS/GMRS radios are a de-facto
standard, the range limitations of the FRS/GMRS standards are eclipsed
by the unlicensed nature of the spectrum\footnote{GMRS has a licensing requirement in the United States, though many
users do not apply for one}. In 2018, these radios have as many as 14 possible <1 km-range (FRS)\citep{FCCFRS}
channels and partially-overlapping 14 3-10 km-range (GMRS)\citep{FCCGMRS,BCLink1,BCLink2}
channels.

In addition to the utility radios have for communication within a
party\citep{SlideSystemicFailure}, radios have the potential to improve
situational awareness and coordination between parties.

\subsubsection{Community channel/party channel}

Some regions, Telluride's Bear Creek in particular, have begun to
define community radio channels \citep{BearCreekEdgerlyISSW,TellurideBackcountryRadio,ConwayRadioProtocol,BCASafetyBlurb,BearCreekShopArticle}.
Both a glance at a map of Bear Creek and the inclusion of two Bear
Creek accidents in this paper suggest that, for Bear Creek locals,
inter-party coordination is essential. While the majority of the documentation
praising the effort has, perhaps necessarily, been from the radio
manufacturer, community radio channels appear to have made a positive
difference in promoting coordination between parties and speeding
rescue efforts. Indeed, the court decision\citep{SanMiguelCountyJudgement}
regarding the Grandfather Couloir accident specifically states that
radios are ``commonly used to facilitate communication between groups
in similar scenarios.''

As party density increases, it may be necessary for backcountry-specific
radio manufacturers to add the ability to monitor a community channel
and an intra-party channel simultaneously, so that intra-party communication
need not obstruct inter-party communication. Such a system could be
confusing in the haste of an incident – good radio discipline on a
party channel is preferable, so long as it remains practical. Steen
and Edgerly\citep{BearCreekEdgerlyISSW} make a similar suggestion
of a the addition of a ``scan'' functionality to future radios.
Aviation radios\citep{RadioGarmin,RadioBendix} have two-channel ``Active''-channel
and ``Standby''-channel functionality, enabling listening to both
and transmitting on one or the other easily.

FRS/GMRS radios, in general, offer the user the reverse feature. Digital
``privacy codes'' can be used define a sub-channel by selective
muting within each channel. Operating the radio tuned to a channel,
but without a ``privacy code'', may allow a listener to hear all
of the traffic on a particular channel. An example: A radio tuned
to ``4-20'' will make audible only those transmissions sent from
other ``4-20'' radios. A radio tuned to ``4'' alone will hear
transmissions from ``4, 4-1, 4-2, ..., 4-20, ... ``. If radio traffic
is not copious, such a setting could provide the wary traveler with
slightly more information about other parties.

\section{Extensions}

The model presented in Section \ref{sec:Estimation} is general, and
should extend to the estimation of rates for other inter-party backcountry
interactions. Krause\citep{DougKrauseEmail} points out that inter-party
rescues may be of similar interest. Inter-party rescues and inter-party
observations are interesting both for understanding the rate at which
inter-party rescues occur and as a proxy for inter-party involvement
rates. As the area, $A_{\text{observation}}$, over which one party
can see or hear an avalanche triggered by another party is much larger
than $A_{\text{avalanche}}$, we should expect the rate at which parties
observe avalanches triggered by others to be much higher than the
rate of inter-party involvements. If $A_{\text{observation}}$ is
substituted for $A_{\text{avalanche}}$, the mathematics should be
the same. The model then predicts that the rate of inter-party avalanche
observations/rescues should be roughly $r_{\text{inter-party observation}}\approx r_{0}A_{\text{observation}}n_{\text{party}}^{2}.$
As avalanche observations and rescues happen far more often than fatal
accidents, there may be an opportunity for new research with more-reliable
statistics.

\section{Conclusion}

We made a first-principles estimation of the scaling of inter-party
avalanche involvements with party-density and found it to be quadratic.
Furthermore, we find that inter-party involvements should become a
significant fraction of all avalanche involvements when the localized
party density approaches one party per avalanche area. This result
is independent of the day's likelihood of triggering avalanches, and
may have utility for avalanche forecasters.

We explored a series of North-American avalanche incidents in order
to show varied ways in which inter-party involvements have occurred.
Moreover, the party-densities at which these incidents occurred are
roughly consistent with the model, with many incidents near $n_{\text{party}}A_{\text{avalanche}}\sim0.1$.

We enumerated and discussed a number of possible strategies for mitigating
inter-party avalanche hazards. Improved awareness, spreading our parties
out over greater area, traveling with others in mind, implementing
regional travel standards, and communicating more-effectively may
all help to limit the number of involvements.

As this work was first drafted in an Autumn-2018 Seattle coffee shop,
customers discussed how their friends encountered ``lines of cars''
at a local hiking trailhead. This aligns with our experience in recent
years, with unprecedented numbers of wilderness travelers at every
trailhead and the disappearance of lonely places in the Cascades.
Inter-party avalanche involvements are coming, if they have not already
arrived. It behooves us, individually and as a community, to develop
strategies to proactively address this emerging hazard.\\
\\
The source code and data needed to generate this document are freely-available\citep{GithubRepository}.

\section{Acknowledgments}

Big thank-yous to Susan Ashlock for her perpetual support for, and
tolerance of, this project; Peg Achterman for helping to bring this
paper back to life; Bruce Jamieson for suggesting the Lizard Range
and Empress Lake incidents; Doug Krause for directing attention to
inter-party avalanche problem-type and inter-party rescues; Jason
Alferness, Dallas Glass, B. J., D.K., and Mark Vesely for assistance
with references; S.A. and John Greendeer Lee for proofreading and
suggestions; and finally, the many avalanche/search-and-rescue organizations
who have not only saved or recovered many of the people described,
but also generated the documentation needed to help us prevent incidents
in the future.

This nights-and-weekends work was indirectly supported by CENPA and
the National Science Foundation (PHY-1607391 and PHY-1912514).

\appendix

\section{Correctly handling $N_{\text{parties}}$\label{sec:Correctly-handling-N}}

Here, we estimate, without approximation with regard to $N_{\text{parties}}$,
the quantities derived in Section \ref{sec:Estimation}.

We begin with the same initial assumption: 

\[
r_{\text{single-party}}=r_{0}\frac{N_{\text{parties}}}{A}
\]

But now, we include the fact that if there are $N_{\text{parties}}$
parties exploring the area $A$, only $N_{\text{parties}}-1$ parties
can be caught in an inter-party avalanche.

\[
r_{\text{interparty}}=r_{\text{single-party}}(N_{\text{parties}}-1)\frac{A_{\text{avalanche}}}{A}
\]

which simplifies to 

\begin{comment}
$r_{\text{inter-party}}=r_{0}\frac{N}{A}(N_{\text{parties}}-1)\frac{A_{\text{avalanche}}}{A}$

$r_{\text{inter-party}}=r_{0}n_{\text{party}}(N_{\text{parties}}-1)\frac{A_{\text{avalanche}}}{A}$
\end{comment}

\[
r_{\text{interparty}}=r_{0}n_{\text{party}}^{2}A_{\text{avalanche}}\frac{N_{\text{parties}}-1}{N_{\text{parties}}}.
\]

Here, we see that the correction we will need is, in general, a factor
of $(N_{\text{parties}}-1)/N_{\text{parties}}$, which is bounded
above by 1 and below by 1/2.

The definition of the total rate remains unchanged

\[
r=r_{\text{single-party}}+r_{\text{interparty}}
\]

and Equation \ref{eq:RelativeIncidentFraction} is modified to

\begin{comment}
$\frac{r_{\text{inter-party}}}{r}=\frac{r_{0}n_{\text{party}}^{2}A_{\text{avalanche}}\frac{N_{\text{parties}}-1}{N_{\text{parties}}}}{r_{0}n_{\text{party}}+r_{0}n_{\text{party}}^{2}A_{\text{avalanche}}\frac{N_{\text{parties}}-1}{N_{\text{parties}}}}$

$\frac{r_{\text{inter-party}}}{r}=\frac{n_{\text{party}}A_{\text{avalanche}}\frac{N_{\text{parties}}-1}{N_{\text{parties}}}}{1+n_{\text{party}}A_{\text{avalanche}}\frac{N_{\text{parties}}-1}{N_{\text{parties}}}}$
\end{comment}

\[
\frac{r_{\text{inter-party}}}{r}=\frac{1}{\frac{1}{n_{\text{party}}A_{\text{avalanche}}\frac{N_{\text{parties}}-1}{N_{\text{parties}}}}+1}.
\]

It is then clear that the true figure of merit is 
\[
n_{\text{party}}A_{\text{avalanche}}\frac{N_{\text{parties}}-1}{N_{\text{parties}}}
\]

and the critical density is

\[
n_{\text{critical}}=\frac{N_{\text{parties}}}{N_{\text{parties}}-1}\frac{1}{A_{\text{avalanche}}}.
\]

This passes the sanity check, as when $N_{\text{parties}}=2$, $n_{\text{critical}}$
is two parties per avalanche area – \emph{all} the parties in the
area have to be in the same avalanche incident in order for an inter-party
involvement to occur. When $N_{\text{parties}}$ is large, no such
coincidence is required.

\section{Estimating the triggering-likelihood $r_{0}$ \label{subsec:Deriving-likelihood}}

It is interesting to discern what underpins the likelihood-of-triggering
$r_{0}$.

Let the area of the day's typical start zones be $A_{\text{start-zone}}$
and the density of start-zones be $n_{\text{start-zones}}\equiv N_{\text{start-zones}}/A$.
In a day's travel, a party will affect the snow on a convoluted strip
of width $w_{\text{party}}$ and length $l$, hence $A_{\text{party}}=lw_{\text{party}}$
or, more-generally, $A_{\text{party}}=w_{\text{party}}s_{\text{party}}t$,
where $s_{\text{party}}$ is the party's speed and $t$ the day's
travel time. For a ski party, $w\sim1$~m, and $s\sim10$~km/d.
The number of slides $N_{\text{slides}}$ triggered in a day is then: 

\[
N_{\text{slides}}=N_{\text{parties}}\frac{A_{\text{party}}}{A}N_{\text{start-zones}}A_{\text{start-zone}}f_{\text{trigger}}
\]

where $f_{\text{trigger}}$ is the number of trigger points per unit
area in the start zones. When avalanche-conditions are `touchy', $f_{\text{trigger}}\sim1$~m$^{-2}$;
when conditions are `stubborn', $f_{\text{trigger }}\ll1$~m$^{-2}$.
$N_{\text{start-zones}}A_{\text{start-zone}}$ is necessarily smaller
than $A$.

This model passes two simple limiting-case checks: If there is but
one square meter of triggerable area, and a single party covers 10~km/d
(0.01~km$^{2}$), and $A=1$~km$^{2}$, we should expect 0.01 slides/day.
If every square meter is a triggerable start-zone, we should expect
to see 10,000 slides/day. 

If we switch to densities ($r_{\text{single-party}}t\equiv N_{\text{slides}}/A$):

\[
r_{\text{single-party}}tA=n_{\text{party}}A_{\text{party}}n_{\text{start-zones}}A_{\text{start-zone}}f_{\text{trigger}}A
\]

\[
r_{\text{single-party}}=w_{\text{party}}s_{\text{party}}f_{\text{trigger}}n_{\text{party}}n_{\text{start-zones}}A_{\text{start-zone}}
\]

where $n_{\text{slides}}$ has units of slides per unit time per unit
area.

For simplicity, we define

\[
r_{0}\equiv w_{\text{party}}s_{\text{party}}f_{\text{trigger}}n_{\text{start-zones}}A_{\text{start-zone}}
\]

so that

\[
r_{\text{single-party}}=r_{0}n_{\text{party}}
\]

as required. $r_{0}$ will depend upon party-type, as $s_{\text{party}}$
varys by at least an order of magnitude between mechanized and non-mechanized
travel.

\subsection{Sensitivity to area estimation\label{subsec:Sensitivity-to-area}}

These estimations are sensitive to the determination of $A$, which
is subjective. If our parties and start-zones are uniformly distributed
(an underlying assumption of the model), these rates are insensitive
to $A$. If our parties tend to congregate near start-zones within
$A$, but the connected terrain is larger, $A$'s determination becomes
challenging.

The crossover, $n_{\text{critical}}$, from single-party involvements
to inter-party involvements is least-sensitive:

\[
n_{\text{party}}A_{\text{avalanche}}=\frac{N_{\text{parties}}A_{\text{avalanche}}}{A}
\]

Absolute-rate estimations are more sensitive, as 

\[
r_{\text{single-party}}=\frac{N_{\text{parties}}N_{\text{start-zones}}}{A^{2}}w_{\text{party}}s_{\text{party}}f_{\text{trigger}}A_{\text{start-zone}}
\]

For these reasons, it will be difficult for quantitative estimates
of these rates and critical densities to be more accurate than a factor
of two in many cases.

\bibliography{SkiDensity}

\begin{thebibliography}{83}%
\makeatletter
\providecommand \@ifxundefined [1]{%
 \@ifx{#1\undefined}
}%
\providecommand \@ifnum [1]{%
 \ifnum #1\expandafter \@firstoftwo
 \else \expandafter \@secondoftwo
 \fi
}%
\providecommand \@ifx [1]{%
 \ifx #1\expandafter \@firstoftwo
 \else \expandafter \@secondoftwo
 \fi
}%
\providecommand \natexlab [1]{#1}%
\providecommand \enquote  [1]{``#1''}%
\providecommand \bibnamefont  [1]{#1}%
\providecommand \bibfnamefont [1]{#1}%
\providecommand \citenamefont [1]{#1}%
\providecommand \href@noop [0]{\@secondoftwo}%
\providecommand \href [0]{\begingroup \@sanitize@url \@href}%
\providecommand \@href[1]{\@@startlink{#1}\@@href}%
\providecommand \@@href[1]{\endgroup#1\@@endlink}%
\providecommand \@sanitize@url [0]{\catcode `\\12\catcode `\$12\catcode
  `\&12\catcode `\#12\catcode `\^12\catcode `\_12\catcode `\%12\relax}%
\providecommand \@@startlink[1]{}%
\providecommand \@@endlink[0]{}%
\providecommand \url  [0]{\begingroup\@sanitize@url \@url }%
\providecommand \@url [1]{\endgroup\@href {#1}{\urlprefix }}%
\providecommand \urlprefix  [0]{URL }%
\providecommand \Eprint [0]{\href }%
\providecommand \doibase [0]{http://dx.doi.org/}%
\providecommand \selectlanguage [0]{\@gobble}%
\providecommand \bibinfo  [0]{\@secondoftwo}%
\providecommand \bibfield  [0]{\@secondoftwo}%
\providecommand \translation [1]{[#1]}%
\providecommand \BibitemOpen [0]{}%
\providecommand \bibitemStop [0]{}%
\providecommand \bibitemNoStop [0]{.\EOS\space}%
\providecommand \EOS [0]{\spacefactor3000\relax}%
\providecommand \BibitemShut  [1]{\csname bibitem#1\endcsname}%
\let\auto@bib@innerbib\@empty
%</preamble>
\bibitem [{\citenamefont {D'Amico}(2016)}]{KendallDecember19}%
  \BibitemOpen
  \bibfield  {author} {\bibinfo {author} {\bibfnamefont {Dennis}\ \bibnamefont
  {D'Amico}},\ }\bibfield  {title} {\enquote {\bibinfo {title} {{K}endall
  {P}eak avalanche fatality, {D}ecember 19, 2015},}\ }\href
  {http://media.nwac.us.s3.amazonaws.com/media/filer_public/9b/ea/9bea699c-d304-48dd-8fff-16c384744e9e/20151219_kendallpkfatalityfinal.pdf}
  {\bibfield  {journal} {\bibinfo  {journal} {Northwest Avalanche Center
  Accident Reports}\ } (\bibinfo {year} {2016})}\BibitemShut {NoStop}%
\bibitem [{\citenamefont {Alliance}(2016)}]{WWAHumans}%
  \BibitemOpen
  \bibfield  {author} {\bibinfo {author} {\bibfnamefont {Winter~Wildlands}\
  \bibnamefont {Alliance}},\ }\href
  {https://winterwildlands.org/wp-content/uploads/2017/04/Economic-Impact-2016.pdf}
  {\enquote {\bibinfo {title} {Human powered snowsports trends and economic
  impacts},}\ } (\bibinfo {year} {2016})\BibitemShut {NoStop}%
\bibitem [{\citenamefont {Merrill}(2019)}]{AvalancheHourEvelynLees}%
  \BibitemOpen
  \bibfield  {author} {\bibinfo {author} {\bibfnamefont {Caleb}\ \bibnamefont
  {Merrill}},\ }\href
  {https://soundcloud.com/user-23585762/the-avalanche-hour-podcast-episode-316-evelyn-lees}
  {\enquote {\bibinfo {title} {Episode 3.16 -- {E}velyn {L}ees},}\ }\bibinfo
  {howpublished} {The Avalanche Hour Podcast} (\bibinfo {year} {2019}),\
  \bibinfo {note} {25:30-27:00}\BibitemShut {NoStop}%
\bibitem [{\citenamefont {Meisenheimer}(2015)}]{MeisenheimerNSAW}%
  \BibitemOpen
  \bibfield  {author} {\bibinfo {author} {\bibfnamefont {Trent}\ \bibnamefont
  {Meisenheimer}},\ }\href@noop {} {\enquote {\bibinfo {title} {How freedom of
  the hills has become anarchy in the backcountry},}\ }\bibinfo {howpublished}
  {Northwest Snow and Avalanche Workshop} (\bibinfo {year} {2015})\BibitemShut
  {NoStop}%
\bibitem [{\citenamefont {Rupf}\ \emph {et~al.}(2019)\citenamefont {Rupf},
  \citenamefont {Haegeli}, \citenamefont {Karlen},\ and\ \citenamefont
  {Wyttenbach}}]{BackcountryCrowdingDynamics}%
  \BibitemOpen
  \bibfield  {author} {\bibinfo {author} {\bibfnamefont {Reto}\ \bibnamefont
  {Rupf}}, \bibinfo {author} {\bibfnamefont {Pascal}\ \bibnamefont {Haegeli}},
  \bibinfo {author} {\bibfnamefont {Barbara}\ \bibnamefont {Karlen}}, \ and\
  \bibinfo {author} {\bibfnamefont {Martin}\ \bibnamefont {Wyttenbach}},\
  }\bibfield  {title} {\enquote {\bibinfo {title} {Does perceived crowding
  cause winter backcountry recreationists to displace?}}\ }\href
  {https://bioone.org/journals/mountain-research-and-development/volume-39/issue-1/MRD-JOURNAL-D-18-00009.1/Does-Perceived-Crowding-Cause-Winter-Backcountry-Recreationists-to-Displace/10.1659/MRD-JOURNAL-D-18-00009.1.full}
  {\bibfield  {journal} {\bibinfo  {journal} {Mountain Research and
  Development}\ }\textbf {\bibinfo {volume} {39}} (\bibinfo {year}
  {2019})}\BibitemShut {NoStop}%
\bibitem [{\citenamefont {Loomis}(2014)}]{BackcountryMagazineCrowding}%
  \BibitemOpen
  \bibfield  {author} {\bibinfo {author} {\bibfnamefont {Molly}\ \bibnamefont
  {Loomis}},\ }\bibfield  {title} {\enquote {\bibinfo {title} {Mountain skills:
  Managing risk and responsibility},}\ }\href
  {https://backcountrymagazine.com/stories/mountain-skills-managing-risk-responsibility/}
  {\bibfield  {journal} {\bibinfo  {journal} {Backcountry Magazine}\ }
  (\bibinfo {year} {2014})}\BibitemShut {NoStop}%
\bibitem [{\citenamefont {McLean}(2015)}]{BackcountryComPowderPolice}%
  \BibitemOpen
  \bibfield  {author} {\bibinfo {author} {\bibfnamefont {Andrew}\ \bibnamefont
  {McLean}},\ }\href
  {https://www.backcountry.com/explore/the-powder-police-keeping-the-backcountry-safe}
  {\enquote {\bibinfo {title} {The {P}owder {P}olice: Keeping the backcountry
  safe},}\ }\bibinfo {howpublished} {Backcountry.com} (\bibinfo {year}
  {2015})\BibitemShut {NoStop}%
\bibitem [{\citenamefont {Krause}(2018)}]{SlidePodcastScene}%
  \BibitemOpen
  \bibfield  {author} {\bibinfo {author} {\bibfnamefont {Doug}\ \bibnamefont
  {Krause}},\ }\href
  {https://soundcloud.com/user-660921194/s2e2-the-scene-of-the-scene} {\enquote
  {\bibinfo {title} {{S2E3}:{T}he {S}cene of the {S}cene},}\ }\bibinfo
  {howpublished} {{S}lide, {T}he {A}valanche {P}odcast} (\bibinfo {year}
  {2018})\BibitemShut {NoStop}%
\bibitem [{\citenamefont {Guy}(Retrieved 2019)}]{CrestedButteEtiquette}%
  \BibitemOpen
  \bibfield  {author} {\bibinfo {author} {\bibfnamefont {Zach}\ \bibnamefont
  {Guy}},\ }\href {http://cbavalanchecenter.org/backcountry-etiquette/}
  {\enquote {\bibinfo {title} {Backcountry etiquette},}\ }\bibinfo
  {howpublished} {Crested Butte Avalanche Center} (\bibinfo {year} {Retrieved
  2019})\BibitemShut {NoStop}%
\bibitem [{\citenamefont {Karkanen}(2014)}]{JumboMountain}%
  \BibitemOpen
  \bibfield  {author} {\bibinfo {author} {\bibfnamefont {Steve}\ \bibnamefont
  {Karkanen}},\ }\href
  {https://missoulaavalanche.org/wp-content/uploads/Mt-Jumbo-avalanche-final1.pdf}
  {\enquote {\bibinfo {title} {{M}ount {J}umbo avalanche accident},}\ }\bibinfo
  {howpublished} {West Central Montana Avalanche Center} (\bibinfo {year}
  {2014})\BibitemShut {NoStop}%
\bibitem [{\citenamefont {Census}(Retrieved 2019)}]{CensusDensity}%
  \BibitemOpen
  \bibfield  {author} {\bibinfo {author} {\bibfnamefont {US}~\bibnamefont
  {Census}},\ }\href
  {https://www.census.gov/quickfacts/fact/map/seattlecitywashington} {\enquote
  {\bibinfo {title} {{U}nited {S}tates {C}ensus {Q}uick{F}acts (2018 population
  estimate 744,955, 83.94 square miles)},}\ } (\bibinfo {year} {Retrieved
  2019})\BibitemShut {NoStop}%
\bibitem [{NWA(2018)}]{NWACAnnualReport2018}%
  \BibitemOpen
  \href {https://www.nwac.us/about/annual-reports/} {\enquote {\bibinfo {title}
  {{NWAC} 2017/2018 {A}nnual {R}eport},}\ }\bibinfo {howpublished} {NWAC.us}
  (\bibinfo {year} {2018})\BibitemShut {NoStop}%
\bibitem [{\citenamefont {Hesser}(2019)}]{HesserTetongBackup}%
  \BibitemOpen
  \bibfield  {author} {\bibinfo {author} {\bibfnamefont {Ted}\ \bibnamefont
  {Hesser}},\ }\href {https://www.instagram.com/p/BvuIVrdlT_l/} {\enquote
  {\bibinfo {title} {Instagram post, {A}pril 1},}\ }\bibinfo {howpublished}
  {Instagram Story and Post} (\bibinfo {year} {2019}),\ \bibinfo {note} {"There
  were 14 other parties attempting the route. I wish I had taken a photo that
  showed all of them bottle-necking... We decided to ski a different line due
  to the objective hazard of parties above and below"}\BibitemShut {NoStop}%
\bibitem [{Cal(2019)}]{Caltopo}%
  \BibitemOpen
  \href@noop {} {\enquote {\bibinfo {title} {Caltopo mapping software},}\
  }\bibinfo {howpublished} {Caltopo.com} (\bibinfo {year} {2019})\BibitemShut
  {NoStop}%
\bibitem [{\citenamefont {Krause}(2019)}]{DougKrauseEmail}%
  \BibitemOpen
  \bibfield  {author} {\bibinfo {author} {\bibfnamefont {Doug}\ \bibnamefont
  {Krause}},\ }\href@noop {} {\enquote {\bibinfo {title} {Personal
  communication},}\ } (\bibinfo {year} {2019})\BibitemShut {NoStop}%
\bibitem [{\citenamefont {Jamieson}\ \emph {et~al.}(2010)\citenamefont
  {Jamieson}, \citenamefont {Haegeli},\ and\ \citenamefont
  {Gauthier}}]{CanadaAvalancheBook}%
  \BibitemOpen
  \bibfield  {author} {\bibinfo {author} {\bibfnamefont {Bruce}\ \bibnamefont
  {Jamieson}}, \bibinfo {author} {\bibfnamefont {Pascal}\ \bibnamefont
  {Haegeli}}, \ and\ \bibinfo {author} {\bibfnamefont {Dave}\ \bibnamefont
  {Gauthier}},\ }\href@noop {} {\emph {\bibinfo {title} {Avalanche Accidents in
  Canada: 1996-2007}}},\ Vol.~\bibinfo {volume} {5}\ (\bibinfo  {publisher}
  {Canadian Avalanche Association},\ \bibinfo {year} {2010})\BibitemShut
  {NoStop}%
\bibitem [{Liz(2001)}]{LizardRangeAvalancheCanada}%
  \BibitemOpen
  \href
  {https://www.avalanche.ca/incidents/454e7f95-2174-4f35-8fa0-7d8a644455a6}
  {\enquote {\bibinfo {title} {Incident summary ({L}izard {R}ange)},}\
  }\bibinfo {howpublished} {Avalanche Canada Website} (\bibinfo {year}
  {2001})\BibitemShut {NoStop}%
\bibitem [{\citenamefont {Frankenfield}(2001)}]{LizardRangeCSAC}%
  \BibitemOpen
  \bibfield  {author} {\bibinfo {author} {\bibfnamefont {Jim}\ \bibnamefont
  {Frankenfield}},\ }\bibfield  {title} {\enquote {\bibinfo {title} {Avalanche
  incident, {F}ebruary 13 2001},}\ }\href
  {http://www.avalanche-center.org/Incidents/2000-01/20010213-Canada.php}
  {\bibfield  {journal} {\bibinfo  {journal} {CyberSpace Avalanche Center}\ }
  (\bibinfo {year} {2001})}\BibitemShut {NoStop}%
\bibitem [{\citenamefont {Walton}(2001)}]{LizardRangeGlobeAndMail}%
  \BibitemOpen
  \bibfield  {author} {\bibinfo {author} {\bibfnamefont {Dawn}\ \bibnamefont
  {Walton}},\ }\bibfield  {title} {\enquote {\bibinfo {title} {Two killed in
  {B}.{C}. avalanche},}\ }\href
  {https://www.theglobeandmail.com/news/national/two-killed-in-bc-avalanche/article1030212/}
  {\bibfield  {journal} {\bibinfo  {journal} {The Globe and Mail}\ } (\bibinfo
  {year} {2001})}\BibitemShut {NoStop}%
\bibitem [{\citenamefont {Morison}(2001)}]{LizardRangeOttawa}%
  \BibitemOpen
  \bibfield  {author} {\bibinfo {author} {\bibfnamefont {Keith}\ \bibnamefont
  {Morison}},\ }\bibfield  {title} {\enquote {\bibinfo {title} {Avalanche kills
  two {S}wedish skiers},}\ }\href
  {https://www.newspapers.com/newspage/466080584/} {\bibfield  {journal}
  {\bibinfo  {journal} {The Ottawa Citizen}\ } (\bibinfo {year}
  {2001})}\BibitemShut {NoStop}%
\bibitem [{Emp(2004)}]{EmpressLakeAvalancheCanada}%
  \BibitemOpen
  \href
  {https://www.avalanche.ca/incidents/f24ca8df-8d11-4fd9-9a3e-1dccdf3f9f54}
  {\enquote {\bibinfo {title} {Incident summary ({M}t. {S}ymons)},}\ }\bibinfo
  {howpublished} {Avalanche Canada Website} (\bibinfo {year}
  {2004})\BibitemShut {NoStop}%
\bibitem [{\citenamefont {Hanke}(2018{\natexlab{a}})}]{HankeNarrative1}%
  \BibitemOpen
  \bibfield  {author} {\bibinfo {author} {\bibfnamefont {Jeremy}\ \bibnamefont
  {Hanke}},\ }\bibfield  {title} {\enquote {\bibinfo {title} {Buried before his
  time – {P}art 1 – {T}he fatal avalanche at {A}rea 51},}\ }\href
  {https://www.mountainlifemedia.ca/2018/12/buried-before-his-time-part-1-the-fatal-avalanche-at-area-51/}
  {\bibfield  {journal} {\bibinfo  {journal} {Mountain Life}\ } (\bibinfo
  {year} {2018}{\natexlab{a}})}\BibitemShut {NoStop}%
\bibitem [{\citenamefont {Hanke}(2018{\natexlab{b}})}]{Hanke2}%
  \BibitemOpen
  \bibfield  {author} {\bibinfo {author} {\bibfnamefont {Jeremy}\ \bibnamefont
  {Hanke}},\ }\bibfield  {title} {\enquote {\bibinfo {title} {Buried before his
  time – {P}art 2 – {L}ife after the avalanche},}\ }\href
  {https://www.mountainlifemedia.ca/2018/12/buried-before-his-time-part-2-life-after-the-avalanche/}
  {\bibfield  {journal} {\bibinfo  {journal} {Mountain Life}\ } (\bibinfo
  {year} {2018}{\natexlab{b}})}\BibitemShut {NoStop}%
\bibitem [{\citenamefont {Hume}(2004)}]{EmpressGlobeAndMail}%
  \BibitemOpen
  \bibfield  {author} {\bibinfo {author} {\bibfnamefont {Mark}\ \bibnamefont
  {Hume}},\ }\bibfield  {title} {\enquote {\bibinfo {title} {One dead after
  group is buried under {B.C.} avalanche},}\ }\href
  {https://www.theglobeandmail.com/news/national/one-dead-after-group-is-buried-under-bc-avalanche/article995885/}
  {\bibfield  {journal} {\bibinfo  {journal} {The Globe and Mail}\ } (\bibinfo
  {year} {2004})}\BibitemShut {NoStop}%
\bibitem [{\citenamefont {Fowler}(2004)}]{EmpressCastanet}%
  \BibitemOpen
  \bibfield  {author} {\bibinfo {author} {\bibfnamefont {David}\ \bibnamefont
  {Fowler}},\ }\bibfield  {title} {\enquote {\bibinfo {title} {Avalanche
  update},}\ }\href {https://www.castanet.net/news/BC/320/Avalanche-Update}
  {\bibfield  {journal} {\bibinfo  {journal} {Castanet}\ } (\bibinfo {year}
  {2004})}\BibitemShut {NoStop}%
\bibitem [{\citenamefont {Hagedorn}(2008{\natexlab{a}})}]{CharlieWyeast}%
  \BibitemOpen
  \bibfield  {author} {\bibinfo {author} {\bibfnamefont {Charlie}\ \bibnamefont
  {Hagedorn}},\ }\href
  {http://www.turns-all-year.com/skiing_snowboarding/trip_reports/index.php?topic=10349.0}
  {\enquote {\bibinfo {title} {June 14, 2008, {W}y'east {F}ace},}\ }\bibinfo
  {howpublished} {Turns-All-Year Forum} (\bibinfo {year}
  {2008}{\natexlab{a}})\BibitemShut {NoStop}%
\bibitem [{\citenamefont {Crockatt}(2008)}]{CrokattHood}%
  \BibitemOpen
  \bibfield  {author} {\bibinfo {author} {\bibfnamefont {Dale}\ \bibnamefont
  {Crockatt}},\ }\href
  {http://www.turns-all-year.com/skiing_snowboarding/trip_reports/index.php?topic=10394.0}
  {\enquote {\bibinfo {title} {June 14 {M}t. {H}ood circumnavigation},}\
  }\bibinfo {howpublished} {Turns-All-Year Forum} (\bibinfo {year}
  {2008})\BibitemShut {NoStop}%
\bibitem [{\citenamefont {Hummel}(2008)}]{HummelCircumnav}%
  \BibitemOpen
  \bibfield  {author} {\bibinfo {author} {\bibfnamefont {Jason}\ \bibnamefont
  {Hummel}},\ }\href
  {http://www.cascadecrusades.org/SkiMountaineering/adams/circumnavigation/circum2008/highorbit2008.htm}
  {\enquote {\bibinfo {title} {{M}ount {A}dams - {C}ircumnavigation {J}une
  13-15th, 2008},}\ }\bibinfo {howpublished} {Cascade Crusades Website}
  (\bibinfo {year} {2008})\BibitemShut {NoStop}%
\bibitem [{\citenamefont {Hagedorn}(2008{\natexlab{b}})}]{CharlieSWChutes}%
  \BibitemOpen
  \bibfield  {author} {\bibinfo {author} {\bibfnamefont {Charlie}\ \bibnamefont
  {Hagedorn}},\ }\href
  {http://www.turns-all-year.com/skiing_snowboarding/trip_reports/index.php?topic=10350.0}
  {\enquote {\bibinfo {title} {June 15, 2008, {A}dams {SW} {C}hutes},}\
  }\bibinfo {howpublished} {Turns-All-Year Forum} (\bibinfo {year}
  {2008}{\natexlab{b}})\BibitemShut {NoStop}%
\bibitem [{Rya(2008)}]{RyanWilsonThread}%
  \BibitemOpen
  \href
  {http://www.turns-all-year.com/skiing_snowboarding/trip_reports/index.php?topic=10337.0}
  {\enquote {\bibinfo {title} {June 14, 2008, {R}ainier, no {N}isqually {C}hute
  fo {F}athers {D}ay},}\ }\bibinfo {howpublished} {Turns-All-Year Forum}
  (\bibinfo {year} {2008})\BibitemShut {NoStop}%
\bibitem [{\citenamefont {Helmstadter}(2008)}]{HelmstaderWilson}%
  \BibitemOpen
  \bibfield  {author} {\bibinfo {author} {\bibfnamefont {Dan}\ \bibnamefont
  {Helmstadter}},\ }\href
  {http://www.turns-all-year.com/skiing_snowboarding/trip_reports/index.php?topic=10371.0}
  {\enquote {\bibinfo {title} {6/14/08, {M}t. {R}ainier, {F}uhrer {T}humb},}\
  }\bibinfo {howpublished} {Turns-All-Year Forum} (\bibinfo {year}
  {2008})\BibitemShut {NoStop}%
\bibitem [{\citenamefont {Richardson}(2010)}]{CodeBlackBoulderMountain}%
  \BibitemOpen
  \bibfield  {author} {\bibinfo {author} {\bibfnamefont {Lisa}\ \bibnamefont
  {Richardson}},\ }\bibfield  {title} {\enquote {\bibinfo {title} {Code
  {B}lack},}\ }\href
  {https://mountainculturegroup.com/wp-content/uploads/2010/11/CodeBlack_KMCW011.pdf}
  {\bibfield  {journal} {\bibinfo  {journal} {Kootenay Mountain Culture}\ }
  (\bibinfo {year} {2010})}\BibitemShut {NoStop}%
\bibitem [{\citenamefont {BC}(2010)}]{BoulderMountainResponseReview}%
  \BibitemOpen
  \bibfield  {author} {\bibinfo {author} {\bibfnamefont {Emergency~Management}\
  \bibnamefont {BC}},\ }\bibfield  {title} {\enquote {\bibinfo {title} {Best
  practices review of the {B}oulder {M}ountain avalanche response},}\ }\href
  {http://revelstokehealthfoundation.com/wp-content/uploads/2016/10/Boulder-Mountain-Avalanche-Response-Review.pdf}
  {\bibfield  {journal} {\bibinfo  {journal} {Emergency Management BC}\ }
  (\bibinfo {year} {2010})}\BibitemShut {NoStop}%
\bibitem [{\citenamefont {Centre}(2010)}]{TurboHillPreliminaryReport}%
  \BibitemOpen
  \bibfield  {author} {\bibinfo {author} {\bibfnamefont {Canadian~Avalanche}\
  \bibnamefont {Centre}},\ }\bibfield  {title} {\enquote {\bibinfo {title}
  {{B}oulder {M}ountain avalanche accident, {M}arch 13, preliminary report},}\
  }\href
  {http://www.avalanche-center.org/Incidents/2009-10/20100313-canada-caa.pdf}
  {\bibfield  {journal} {\bibinfo  {journal} {Canadian Avalanche Centre}\ }
  (\bibinfo {year} {2010})}\BibitemShut {NoStop}%
\bibitem [{\citenamefont
  {Canada}(2010)}]{ParksCanadaBoulderMountainPresentation}%
  \BibitemOpen
  \bibfield  {author} {\bibinfo {author} {\bibfnamefont {Parks}\ \bibnamefont
  {Canada}},\ }\href
  {http://www.alpine-rescue.org/ikar-cisa/documents/2010/ikar20101020000602.pdf}
  {\enquote {\bibinfo {title} {Avalanche rescue, {B}oulder {M}ountain,
  {S}elkirk {M}ountains, {C}anada, {M}arch 13-14, 2010},}\ }\bibinfo
  {howpublished} {Presentation} (\bibinfo {year} {2010})\BibitemShut {NoStop}%
\bibitem [{Eag(2010{\natexlab{a}})}]{EaglePassAvalancheCanada}%
  \BibitemOpen
  \href
  {https://www.avalanche.ca/incidents/5077f4cd-6779-49eb-8848-08c6d6bacde8}
  {\enquote {\bibinfo {title} {Incident summary ({E}agle {P}ass {M}tn)},}\
  }\bibinfo {howpublished} {Avalanche Canada Website} (\bibinfo {year}
  {2010}{\natexlab{a}})\BibitemShut {NoStop}%
\bibitem [{Eag(2010{\natexlab{b}})}]{EaglePassCSAC}%
  \BibitemOpen
  \href
  {https://www.avalanche-center.org/Incidents/2009-10/20100319-Canada.php}
  {\enquote {\bibinfo {title} {Avalanche {I}nvolvement {R}eport ({E}agle
  {P}ass)},}\ }\bibinfo {howpublished} {CyberSpace Avalanche Center} (\bibinfo
  {year} {2010}{\natexlab{b}})\BibitemShut {NoStop}%
\bibitem [{\citenamefont {Fong}\ and\ \citenamefont
  {Press}(2010)}]{EaglePassNewsArticle1}%
  \BibitemOpen
  \bibfield  {author} {\bibinfo {author} {\bibfnamefont {Petti}\ \bibnamefont
  {Fong}}\ and\ \bibinfo {author} {\bibfnamefont {The~Canadian}\ \bibnamefont
  {Press}},\ }\bibfield  {title} {\enquote {\bibinfo {title} {No one else
  feared buried after new {B.C.} avalanche kills snowmobiler},}\ }\href
  {https://www.thestar.com/news/canada/2010/03/20/no_one_else_feared_buried_after_new_bc_avalanche_kills_snowmobiler.html}
  {\bibfield  {journal} {\bibinfo  {journal} {The Star}\ } (\bibinfo {year}
  {2010})}\BibitemShut {NoStop}%
\bibitem [{\citenamefont {Staff}(2010)}]{EaglePassNewsArticle2}%
  \BibitemOpen
  \bibfield  {author} {\bibinfo {author} {\bibfnamefont {CTV}\ \bibnamefont
  {Staff}},\ }\bibfield  {title} {\enquote {\bibinfo {title} {Another {B.C}.
  avalanche kills at least one snowmobiler},}\ }\href
  {https://www.ctvnews.ca/another-b-c-avalanche-kills-at-least-one-snowmobiler-1.493979}
  {\bibfield  {journal} {\bibinfo  {journal} {CTV News}\ } (\bibinfo {year}
  {2010})}\BibitemShut {NoStop}%
\bibitem [{\citenamefont {Kobernik}\ and\ \citenamefont
  {Hardesty}(2011)}]{GadValleyUACFrenzy}%
  \BibitemOpen
  \bibfield  {author} {\bibinfo {author} {\bibfnamefont {Brett}\ \bibnamefont
  {Kobernik}}\ and\ \bibinfo {author} {\bibfnamefont {Andrew}\ \bibnamefont
  {Hardesty}},\ }\bibfield  {title} {\enquote {\bibinfo {title} {Accident:
  {G}ad {V}alley},}\ }\href {https://utahavalanchecenter.org/avalanche/17964}
  {\bibfield  {journal} {\bibinfo  {journal} {Utah Avalanche Center Accident
  Reports}\ } (\bibinfo {year} {2011})}\BibitemShut {NoStop}%
\bibitem [{Tay(2012)}]{TaylorMountainJHAvalanche}%
  \BibitemOpen
  \href {http://jhavalanche.org/eventDetail/event/2621} {\enquote {\bibinfo
  {title} {{S}outh face of {T}aylor {M}ountain zone: 106},}\ }\bibinfo
  {howpublished} {Bridger Teton Avalanche Center} (\bibinfo {year}
  {2012})\BibitemShut {NoStop}%
\bibitem [{\citenamefont {Wolfe}(2012)}]{TaylorMountainTARMusings}%
  \BibitemOpen
  \bibfield  {author} {\bibinfo {author} {\bibfnamefont {Lynne}\ \bibnamefont
  {Wolfe}},\ }\bibfield  {title} {\enquote {\bibinfo {title} {Taylor
  musings},}\ }\href
  {https://www.americanavalancheassociation.org/s/TAR3004_ALL_LoRes.pdf}
  {\bibfield  {journal} {\bibinfo  {journal} {The Avalanche Review}\ }\textbf
  {\bibinfo {volume} {30}},\ \bibinfo {pages} {14--16} (\bibinfo {year}
  {2012})}\BibitemShut {NoStop}%
\bibitem [{\citenamefont {Romeo}(2012)}]{TaylorMountainTetonAT}%
  \BibitemOpen
  \bibfield  {author} {\bibinfo {author} {\bibfnamefont {Steve}\ \bibnamefont
  {Romeo}},\ }\href
  {http://www.tetonat.com/2012/01/24/taylor-mountain-avalanche/} {\enquote
  {\bibinfo {title} {{T}aylor {M}ountain avalanche},}\ }\bibinfo {howpublished}
  {TetonAT.com} (\bibinfo {year} {2012})\BibitemShut {NoStop}%
\bibitem [{\citenamefont {Horne}(2012)}]{TaylorMountainTetonValleyNews}%
  \BibitemOpen
  \bibfield  {author} {\bibinfo {author} {\bibfnamefont {Rachael}\ \bibnamefont
  {Horne}},\ }\bibfield  {title} {\enquote {\bibinfo {title} {Massive slide on
  {T}aylor highlights avalanche danger},}\ }\href
  {https://www.tetonvalleynews.net/news/massive-slide-on-taylor-highlights-avalanche-danger/article_9e5a2fdc-47b2-11e1-9ce0-001871e3ce6c.html}
  {\bibfield  {journal} {\bibinfo  {journal} {Teton Valley News}\ } (\bibinfo
  {year} {2012})}\BibitemShut {NoStop}%
\bibitem [{\citenamefont {Hagedorn}(2017)}]{KendallTrap}%
  \BibitemOpen
  \bibfield  {author} {\bibinfo {author} {\bibfnamefont {Charlie}\ \bibnamefont
  {Hagedorn}},\ }\href {http://www.kendallpeak.org/} {\enquote {\bibinfo
  {title} {{T}he {K}endall {T}rap},}\ }\bibinfo {howpublished}
  {KendallPeak.org} (\bibinfo {year} {2017})\BibitemShut {NoStop}%
\bibitem [{\citenamefont {Grey}(2017)}]{AvalancheCrestBigLines}%
  \BibitemOpen
  \bibfield  {author} {\bibinfo {author} {\bibfnamefont {Tim}\ \bibnamefont
  {Grey}},\ }\href
  {http://biglines.com/conditions/avalanche/rogers-pass-avalanche-february-142016-dan-michalchuk/}
  {\enquote {\bibinfo {title} {{R}ogers {P}ass avalanche, {F}ebruary 14/2016 --
  {D}an {M}ichalchuk interview},}\ }\bibinfo {howpublished} {BigLines.com}
  (\bibinfo {year} {2017})\BibitemShut {NoStop}%
\bibitem [{\citenamefont {Tomlinson}(2017)}]{AvalancheCrestRogersPassScared}%
  \BibitemOpen
  \bibfield  {author} {\bibinfo {author} {\bibfnamefont {Phil}\ \bibnamefont
  {Tomlinson}},\ }\href
  {https://mountainwagon.com/the-blog/the-rogers-pass-trip-for-people-scared-of-rogers-pass}
  {\enquote {\bibinfo {title} {The {R}ogers {P}ass trip for people scared of
  {R}ogers {P}ass},}\ }\bibinfo {howpublished} {MountainWagon.com} (\bibinfo
  {year} {2017})\BibitemShut {NoStop}%
\bibitem [{\citenamefont {Michalchuk}(2017)}]{AvalancheCrestAvalancheVideo}%
  \BibitemOpen
  \bibfield  {author} {\bibinfo {author} {\bibfnamefont {Dan}\ \bibnamefont
  {Michalchuk}},\ }\href {https://vimeo.com/204294001} {\enquote {\bibinfo
  {title} {{R}ogers {P}ass: {A}valanche {C}rest avalanche {F}ebruary
  14/2016},}\ }\bibinfo {howpublished} {Vimeo.com} (\bibinfo {year}
  {2017})\BibitemShut {NoStop}%
\bibitem [{\citenamefont
  {Davis}(2016)}]{GrandfatherCouloirTwoPartyInteraction}%
  \BibitemOpen
  \bibfield  {author} {\bibinfo {author} {\bibfnamefont {Jeff}\ \bibnamefont
  {Davis}},\ }\bibfield  {title} {\enquote {\bibinfo {title} {2016/04/03 -
  {C}olorado - {G}randfather {C}ouloir, {B}ear {C}reek drainage},}\ }\href
  {https://avalanche.state.co.us/caic/acc/acc_report.php?accfm=inv&acc_id=611}
  {\bibfield  {journal} {\bibinfo  {journal} {Colorado Avalanche Information
  Center Accident Database}\ } (\bibinfo {year} {2016})}\BibitemShut {NoStop}%
\bibitem [{\citenamefont {Jackson}(2017)}]{SanMiguelCountyJudgement}%
  \BibitemOpen
  \bibfield  {author} {\bibinfo {author} {\bibfnamefont {D.~Cory}\ \bibnamefont
  {Jackson}},\ }\bibfield  {title} {\enquote {\bibinfo {title} {Order of
  {J}udgement, {C}ase {N}umber 2016{S}26},}\ }\href@noop {} {\bibfield
  {journal} {\bibinfo  {journal} {San Miguel County Court}\ } (\bibinfo {year}
  {2017})}\BibitemShut {NoStop}%
\bibitem [{\citenamefont {Moss}(2018)}]{GrandfatherLegalBlog}%
  \BibitemOpen
  \bibfield  {author} {\bibinfo {author} {\bibfnamefont {James~H.}\
  \bibnamefont {Moss}},\ }\href
  {https://recreation-law.com/2018/01/08/backcountry-skier-sues-in-small-claims-court-in-san-miguel-county-colorado-for-injuries-she-received-when-a-backcountry-snowboarder-triggered-an-avalanche-that-injured-her/}
  {\enquote {\bibinfo {title} {Backcountry skier sues in {S}mall {C}laims
  {C}ourt in {S}an {M}iguel {C}ounty {C}olorado for injuries she received when
  a backcountry snowboarder triggered an avalanche that injured her.}}\
  }\bibinfo {howpublished} {Recreation-Law.com} (\bibinfo {year}
  {2018})\BibitemShut {NoStop}%
\bibitem [{\citenamefont {NWAC}(2017)}]{1617AccidentSummaries}%
  \BibitemOpen
  \bibfield  {author} {\bibinfo {author} {\bibnamefont {NWAC}},\ }\bibfield
  {title} {\enquote {\bibinfo {title} {2016/2017 {A}ccident summaries},}\
  }\href {https://www.nwac.us/accidents/accident-reports/} {\bibfield
  {journal} {\bibinfo  {journal} {Northwest Avalanche Center}\ } (\bibinfo
  {year} {2017})}\BibitemShut {NoStop}%
\bibitem [{\citenamefont {D'Amico}\ and\ \citenamefont
  {Hahn}(2017)}]{HawkinsMountain}%
  \BibitemOpen
  \bibfield  {author} {\bibinfo {author} {\bibfnamefont {Dennis}\ \bibnamefont
  {D'Amico}}\ and\ \bibinfo {author} {\bibfnamefont {Robert}\ \bibnamefont
  {Hahn}},\ }\bibfield  {title} {\enquote {\bibinfo {title} {{H}awkins
  {M}ountain avalanche fatality, {M}arch 4th, 2017},}\ }\href
  {http://media.nwac.us.s3.amazonaws.com/media/filer_public/84/7f/847f1404-1561-4d3c-bcdb-2fa3728d41d4/20170304_hawkinsmountain_fatality_draft.pdf}
  {\bibfield  {journal} {\bibinfo  {journal} {Northwest Avalanche Center
  Accident Reports}\ } (\bibinfo {year} {2017})}\BibitemShut {NoStop}%
\bibitem [{Sno(2017)}]{SnowestHawkins}%
  \BibitemOpen
  \href {https://www.snowest.com/forum/showthread.php?t=431228} {\enquote
  {\bibinfo {title} {{S}nowest {H}awkins {M}ountain accident thread},}\ }
  (\bibinfo {year} {2017})\BibitemShut {NoStop}%
\bibitem [{\citenamefont {Blevins}(2019)}]{TemptationSlideArticle}%
  \BibitemOpen
  \bibfield  {author} {\bibinfo {author} {\bibfnamefont {Jason}\ \bibnamefont
  {Blevins}},\ }\bibfield  {title} {\enquote {\bibinfo {title} {A ducked rope
  line. {A} massive avalanche: Detailed report on deadly {T}elluride slide
  comes as sheriff's probe continues},}\ }\href
  {https://coloradosun.com/2019/03/22/salvador-garcia-atance-avalanche-report-telluride/}
  {\bibfield  {journal} {\bibinfo  {journal} {Colorado Sun}\ } (\bibinfo {year}
  {2019})}\BibitemShut {NoStop}%
\bibitem [{\citenamefont {Davis}\ \emph {et~al.}(2019)\citenamefont {Davis},
  \citenamefont {Logan}, \citenamefont {Greene},\ and\ \citenamefont
  {Bilbrey}}]{TemptationCAICAccidentReport}%
  \BibitemOpen
  \bibfield  {author} {\bibinfo {author} {\bibfnamefont {Jeff}\ \bibnamefont
  {Davis}}, \bibinfo {author} {\bibfnamefont {Spencer}\ \bibnamefont {Logan}},
  \bibinfo {author} {\bibfnamefont {Ethan}\ \bibnamefont {Greene}}, \ and\
  \bibinfo {author} {\bibfnamefont {Chris}\ \bibnamefont {Bilbrey}},\
  }\bibfield  {title} {\enquote {\bibinfo {title} {2019/02/19 - {C}olorado -
  {T}emptation avalanche path, {B}ear {C}reek, south of {T}elluride},}\ }\href
  {https://avalanche.state.co.us/caic/acc/acc_report.php?accfm=inv&acc_id=705}
  {\bibfield  {journal} {\bibinfo  {journal} {Colorado Avalanche Information
  Center Accident Reports}\ } (\bibinfo {year} {2019})}\BibitemShut {NoStop}%
\bibitem [{\citenamefont {Greene}(2010)}]{SWAG}%
  \BibitemOpen
  \bibinfo {editor} {\bibfnamefont {Ethan}\ \bibnamefont {Greene}},\ ed.,\
  \href@noop {} {\emph {\bibinfo {title} {Snow, Weather, and Avalanches:
  Observation Guidelines for Avalanche Programs in the {U}nited {S}tates}}}\
  (\bibinfo  {publisher} {American Avalanche Association},\ \bibinfo {year}
  {2010})\BibitemShut {NoStop}%
\bibitem [{\citenamefont {Scott}(2008)}]{MousetrapAsulkanGuide}%
  \BibitemOpen
  \bibfield  {author} {\bibinfo {author} {\bibfnamefont {Chic}\ \bibnamefont
  {Scott}},\ }\href@noop {} {\emph {\bibinfo {title} {Alpine Ski Tours in the
  {C}olumbia {M}ountains: Summits \& Icefields}}},\ \bibinfo {edition} {second
  edition, third printing}\ ed.\ (\bibinfo  {publisher} {Rocky Mountain
  Books},\ \bibinfo {year} {2008})\ \bibinfo {note} {pages 34-36}\BibitemShut
  {NoStop}%
\bibitem [{\citenamefont {Davis}(2019)}]{NWACRunList}%
  \BibitemOpen
  \bibfield  {author} {\bibinfo {author} {\bibfnamefont {Justin}\ \bibnamefont
  {Davis}},\ }\href
  {https://medium.com/northwest-avalanche-center-nwac/run-list-a-framework-for-tour-planning-2cbcc5ddbe51}
  {\enquote {\bibinfo {title} {Run list: A framework for tour planning},}\
  }\bibinfo {howpublished} {Northwest Avalanche Center -- Medium} (\bibinfo
  {year} {2019})\BibitemShut {NoStop}%
\bibitem [{\citenamefont {Israelson}(2016)}]{RunListIsraelson}%
  \BibitemOpen
  \bibfield  {author} {\bibinfo {author} {\bibfnamefont {Clair}\ \bibnamefont
  {Israelson}},\ }\bibfield  {title} {\enquote {\bibinfo {title} {A suggested
  conceptual model for daily terrain use decisions at {N}orthern {E}scape
  {H}eli-{S}kiing},}\ }\href
  {https://issuu.com/theavalanchejournal/docs/the_avalanche_journal_volume_110}
  {\bibfield  {journal} {\bibinfo  {journal} {The Avalanche Journal}\ }\textbf
  {\bibinfo {volume} {110}},\ \bibinfo {pages} {39--45} (\bibinfo {year}
  {2016})}\BibitemShut {NoStop}%
\bibitem [{FAA(2017)}]{FAAApproaches}%
  \BibitemOpen
  \href
  {https://www.faa.gov/regulations_policies/handbooks_manuals/aviation/instrument_procedures_handbook/}
  {\enquote {\bibinfo {title} {Instrument {P}rocedures {H}andbook:
  {A}pproaches},}\ }\bibinfo {howpublished} {FAA} (\bibinfo {year}
  {2017})\BibitemShut {NoStop}%
\bibitem [{Blu(2013{\natexlab{a}})}]{BlueberryBagleyTAY}%
  \BibitemOpen
  \href
  {http://www.turns-all-year.com/skiing_snowboarding/trip_reports/index.php?topic=29641.0}
  {\enquote {\bibinfo {title} {{B}agley {L}akes/ {A}rtist {P}oint {N}ov 16},}\
  }\bibinfo {howpublished} {Turns-All-Year Forum} (\bibinfo {year}
  {2013}{\natexlab{a}})\BibitemShut {NoStop}%
\bibitem [{Blu(2013{\natexlab{b}})}]{BlueberryDecemberHop}%
  \BibitemOpen
  \href
  {http://www.turns-all-year.com/skiing_snowboarding/trip_reports/index.php?topic=29855.0}
  {\enquote {\bibinfo {title} {Dec. 3, {B}agley {L}akes (continuation of {N}ov.
  16....)},}\ }\bibinfo {howpublished} {Turns-All-Year Forum} (\bibinfo {year}
  {2013}{\natexlab{b}})\BibitemShut {NoStop}%
\bibitem [{\citenamefont {Ü}(2015)}]{BlueberrySeattleSkintrack}%
  \BibitemOpen
  \bibfield  {author} {\bibinfo {author} {\bibfnamefont {Adam}\ \bibnamefont
  {Ü}},\ }\href
  {http://www.turns-all-year.com/skiing_snowboarding/trip_reports/index.php?topic=33188.0}
  {\enquote {\bibinfo {title} {The "{S}eattle {S}kintrack" on {T}able
  {M}ountain},}\ }\bibinfo {howpublished} {Turns-All-Year Forum} (\bibinfo
  {year} {2015})\BibitemShut {NoStop}%
\bibitem [{Blu(2012)}]{BlueberryAprilFools}%
  \BibitemOpen
  \href
  {http://www.turns-all-year.com/skiing_snowboarding/trip_reports/index.php?topic=24400.0}
  {\enquote {\bibinfo {title} {April fools on {T}able {M}ountain},}\ }\bibinfo
  {howpublished} {Turns-All-Year Forum} (\bibinfo {year} {2012})\BibitemShut
  {NoStop}%
\bibitem [{\citenamefont {B}(2012)}]{BlueberryAprilFoolsVideo}%
  \BibitemOpen
  \bibfield  {author} {\bibinfo {author} {\bibfnamefont {Chuck}\ \bibnamefont
  {B}},\ }\href {https://vimeo.com/39808079} {\enquote {\bibinfo {title} {April
  1 on {T}able {M}ountain},}\ }\bibinfo {howpublished} {Vimeo} (\bibinfo {year}
  {2012})\BibitemShut {NoStop}%
\bibitem [{\citenamefont {Fortier}(2000)}]{PhilEnigma}%
  \BibitemOpen
  \bibfield  {author} {\bibinfo {author} {\bibfnamefont {Phil}\ \bibnamefont
  {Fortier}},\ }\href {http://mtnphil.com/enigma2/enigma2.html} {\enquote
  {\bibinfo {title} {{S}noqualmie {M}ountain, {N}orthwest {F}ace},}\ }
  (\bibinfo {year} {2000})\BibitemShut {NoStop}%
\bibitem [{\citenamefont {Hummel}(2004)}]{HummelSlot}%
  \BibitemOpen
  \bibfield  {author} {\bibinfo {author} {\bibfnamefont {Jason}\ \bibnamefont
  {Hummel}},\ }\href
  {http://www.cascadecrusades.org/SkiMountaineering/snoqualmiemtn/slotcouloir/slotcrooked2004.htm}
  {\enquote {\bibinfo {title} {Mount {S}noqualmie, 9415 feet},}\ }\bibinfo
  {howpublished} {Cascade Crusades Website} (\bibinfo {year}
  {2004})\BibitemShut {NoStop}%
\bibitem [{\citenamefont {NWAC}(2018)}]{NWACUphillPanel}%
  \BibitemOpen
  \bibfield  {author} {\bibinfo {author} {\bibnamefont {NWAC}},\ }\href
  {https://www.youtube.com/watch?v=JgSXV-551rw} {\enquote {\bibinfo {title}
  {{NSAW} 2018 - {P}anel {D}iscussion: {S}ki {A}rea {U}phill \& {B}oundary
  {T}ravel},}\ }\bibinfo {howpublished} {Youtube} (\bibinfo {year}
  {2018})\BibitemShut {NoStop}%
\bibitem [{Uph(2019)}]{UphillPolicies}%
  \BibitemOpen
  \href {https://ussma.org/resort-uphill-policies/} {\enquote {\bibinfo {title}
  {Resort uphill policies},}\ }\bibinfo {howpublished} {United States Ski
  Mountaineering Association Website} (\bibinfo {year} {2019})\BibitemShut
  {NoStop}%
\bibitem [{FCC(Retrieved 2019{\natexlab{a}})}]{FCCFRS}%
  \BibitemOpen
  \href
  {https://www.fcc.gov/wireless/bureau-divisions/mobility-division/family-radio-service-frs}
  {\enquote {\bibinfo {title} {Family {R}adio {S}ervice ({FRS})},}\ }\bibinfo
  {howpublished} {Federal Communications Commission} (\bibinfo {year}
  {Retrieved 2019}{\natexlab{a}})\BibitemShut {NoStop}%
\bibitem [{FCC(Retrieved 2019{\natexlab{b}})}]{FCCGMRS}%
  \BibitemOpen
  \href {https://www.fcc.gov/general-mobile-radio-service-gmrs} {\enquote
  {\bibinfo {title} {{G}eneral {M}obile {R}adio {S}ervice ({GMRS})},}\
  }\bibinfo {howpublished} {Federal Communications Commission} (\bibinfo {year}
  {Retrieved 2019}{\natexlab{b}})\BibitemShut {NoStop}%
\bibitem [{BCL(2015)}]{BCLink1}%
  \BibitemOpen
  \href
  {http://www.backcountryaccess.com/wp-content/uploads/2015/08/BCA-product-manual-BC-Link-two-way-radios-english.pdf}
  {\emph {\bibinfo {title} {BC Link Owner's Manual}}},\ \bibinfo {organization}
  {Backcountry Access} (\bibinfo {year} {2015})\BibitemShut {NoStop}%
\bibitem [{BCL(2019)}]{BCLink2}%
  \BibitemOpen
  \href
  {https://www.backcountryaccess.com/wp-content/uploads/2019/01/BCLink2.0_manual_v11.pdf}
  {\emph {\bibinfo {title} {BC Link 2.0 Owner's Manual}}},\ \bibinfo
  {organization} {Backcountry Access},\ \bibinfo {edition} {11th}\ ed.
  (\bibinfo {year} {2019})\BibitemShut {NoStop}%
\bibitem [{\citenamefont {Krause}(2017)}]{SlideSystemicFailure}%
  \BibitemOpen
  \bibfield  {author} {\bibinfo {author} {\bibfnamefont {Doug}\ \bibnamefont
  {Krause}},\ }\href
  {https://soundcloud.com/user-660921194/s1e10-systemic-failure} {\enquote
  {\bibinfo {title} {{S1E10}: {S}ystemic failure},}\ }\bibinfo {howpublished}
  {Slide, {T}he {A}valanche {P}odcast} (\bibinfo {year} {2017})\BibitemShut
  {NoStop}%
\bibitem [{\citenamefont {Steen}\ and\ \citenamefont
  {Edgerly}(2016)}]{BearCreekEdgerlyISSW}%
  \BibitemOpen
  \bibfield  {author} {\bibinfo {author} {\bibfnamefont {Matt}\ \bibnamefont
  {Steen}}\ and\ \bibinfo {author} {\bibfnamefont {Bruce}\ \bibnamefont
  {Edgerly}},\ }\bibfield  {title} {\enquote {\bibinfo {title} {Utilizing
  common radio channels in high-use avalanche terrain},}\ }\href
  {http://arc.lib.montana.edu/snow-science/item/2387} {\bibfield  {journal}
  {\bibinfo  {journal} {Proceedings of the International Snow Science Workshop,
  Breckenridge, Colorado}\ } (\bibinfo {year} {2016})}\BibitemShut {NoStop}%
\bibitem [{Tel(2016)}]{TellurideBackcountryRadio}%
  \BibitemOpen
  \href
  {https://www.telluridemountainclub.org/2016-2017-backcountry-radio-program/}
  {\enquote {\bibinfo {title} {Telluride backcountry radio information},}\
  }\bibinfo {howpublished} {Telluride Mountain Club} (\bibinfo {year}
  {2016})\BibitemShut {NoStop}%
\bibitem [{Con(2016)}]{ConwayRadioProtocol}%
  \BibitemOpen
  \href
  {https://backcountryaccess.com/benefits-protocols-using-two-way-radios-backcountry/}
  {\enquote {\bibinfo {title} {Roger that! "{S}arge" {C}onway on backcountry
  two-way radio protocol},}\ }\bibinfo {howpublished} {Backcountry Access
  Website} (\bibinfo {year} {2016})\BibitemShut {NoStop}%
\bibitem [{BCA(2016)}]{BCASafetyBlurb}%
  \BibitemOpen
  \href
  {https://backcountryaccess.com/the-right-frequency-how-radios-save-lives-in-telluride/}
  {\enquote {\bibinfo {title} {The right frequency: {H}ow radios save lives in
  {T}elluride},}\ }\bibinfo {howpublished} {Backcountry Access Website}
  (\bibinfo {year} {2016})\BibitemShut {NoStop}%
\bibitem [{\citenamefont {Story}(2015)}]{BearCreekShopArticle}%
  \BibitemOpen
  \bibfield  {author} {\bibinfo {author} {\bibfnamefont {Rob}\ \bibnamefont
  {Story}},\ }\href
  {https://www.telluridenews.com/news/article_537a0278-9f95-11e5-bb80-b342a2bb57ce.html}
  {\enquote {\bibinfo {title} {Radios aid backcountry communication: Snow
  safety experts push new models},}\ }\bibinfo {howpublished} {Telluride Daily
  Planet} (\bibinfo {year} {2015})\BibitemShut {NoStop}%
\bibitem [{Rad(2019{\natexlab{a}})}]{RadioGarmin}%
  \BibitemOpen
  \href {https://buy.garmin.com/en-US/US/p/140495} {\enquote {\bibinfo {title}
  {{GTR} 200},}\ }\bibinfo {howpublished} {Garmin Avionics Website} (\bibinfo
  {year} {2019}{\natexlab{a}})\BibitemShut {NoStop}%
\bibitem [{Rad(2019{\natexlab{b}})}]{RadioBendix}%
  \BibitemOpen
  \href {https://www.bendixking.com/en/products/productitems/ky-196a} {\enquote
  {\bibinfo {title} {{KY} 196{A}},}\ }\bibinfo {howpublished} {BendixKing
  Website} (\bibinfo {year} {2019}{\natexlab{b}})\BibitemShut {NoStop}%
\bibitem [{\citenamefont {Hagedorn}(2019)}]{GithubRepository}%
  \BibitemOpen
  \bibfield  {author} {\bibinfo {author} {\bibfnamefont {Charles}\ \bibnamefont
  {Hagedorn}},\ }\href {https://github.com/4kbt/InterPartyAvalancheEstimation}
  {\enquote {\bibinfo {title} {{I}nter{P}arty{A}valanche{E}stimation},}\
  }\bibinfo {howpublished} {Git{H}ub} (\bibinfo {year} {2019})\BibitemShut
  {NoStop}%
\end{thebibliography}%

\end{document}